\documentclass[%
twocolumn,
superscriptaddress,
 amsmath,amssymb,
 aps, physrev,
]{revtex4-2}

\usepackage{graphicx}
\usepackage{dcolumn}
\usepackage{bm}
\usepackage{float}
\usepackage{booktabs}

\usepackage{amsmath,amssymb,amsfonts}%
\usepackage{siunitx}

\newcommand{\MM}{Sec.~Materials and Methods}
\newcommand{\SuppInf}{Supplementary Information}

\renewcommand{\vec}[1]{\mathbf{#1}}

\begin{document}

\preprint{APS/123-QED}

\title{\textbf{Probing the Microscopic Origin of Toughness in Multiple Polymer Networks} 
}%

\author{Nicholas H. P. Orr}
\affiliation{Laboratoire Charles Coulomb (L2C), University of Montpellier, CNRS, Montpellier 34095, France}
\affiliation{CPCV, Department of Chemistry, École Normale Supérieure
PSL University, Sorbonne Université, CNRS
Paris 75005, France.}
\author{Magali Le Goff}
\affiliation{LIPhy, University Grenoble-Alpes, CNRS, Grenoble 38000, France}
\affiliation{Institut für Theoretische Physik, Universität Innsbruck, Technikerstraße 21A, A-6020 Innsbruck, Austria}
\author{Burebi Yiming}
\affiliation{Sciences et Ingénierie de la Matière Molle, CNRS UMR 7615, École supérieure de physique et de chimie industrielles de la Ville de Paris, Sorbonne Université, Paris Sciences et Lettres Université, Paris 75005, France}
\author{Jean-Louis Barrat}
\affiliation{LIPhy, University Grenoble-Alpes, CNRS, Grenoble 38000, France}
\author{Mehdi Bouzid}
\affiliation{3SR, University Grenoble-Alpes, CNRS, 38000, Grenoble INP, Grenoble 38000, Grenoble, France}
\author{Laurence Ramos}
\affiliation{Laboratoire Charles Coulomb (L2C), University of Montpellier, CNRS, Montpellier 34095, France}
\author{Costantino Creton}
\affiliation{Sciences et Ingénierie de la Matière Molle, CNRS UMR 7615, École supérieure de physique et de chimie industrielles de la Ville de Paris, Sorbonne Université, Paris Sciences et Lettres Université, Paris 75005, France}
\author{Kirsten Martens}
\affiliation{LIPhy, University Grenoble-Alpes, CNRS, Grenoble 38000, France}
\author{Luca Cipelletti}
\affiliation{Laboratoire Charles Coulomb (L2C), University of Montpellier, CNRS, Montpellier 34095, France}
\affiliation{Institut Universitaire de France, Paris 75005, France}

\date{\today}

\begin{abstract}
Multiple polymer networks, such as double-network elastomers comprising a sacrificial and a matrix network, exhibit exceptional mechanical resilience, commonly attributed to the formation of an extended damage zone before a crack can grow. However, the microscopic mechanisms underlying their toughness remain poorly understood. Here, we combine advanced light scattering methods and molecular dynamics simulations to explore the microscopic relaxation dynamics and stress redistribution at the polymer strand scale of single-network and double-network elastomers under uni-axial loading. Dynamic light scattering experiments show that microscopic rearrangements and bond breaking events are localized near the crack tip in single networks, readily causing the crack to advance. In contrast, double networks exhibit delocalized microscopic rearrangements well ahead of and not directly correlated with crack propagation, enabling the dissipation of energy over broader regions and timescales. Numerical simulations of the damage zone show that bond breaking in the matrix network of double networks leads to widespread stress redistribution, mitigating catastrophic damage localization. This enhanced ability to redistribute stress in a non-local manner allows a much larger extension before localized macroscopic failure occurs, explaining the superior toughness of double networks. Our findings identify early, delocalized bond breaking events combined with more efficient dissipation pathways through enhanced microscopic rearrangements as the key microscopic mechanisms responsible for the outstanding toughness and extensibility of multiple elastomer networks.
\end{abstract}

\maketitle


Understanding how materials fail is essential for predicting their lifetime and guiding the design of resilient systems. Broadly, failure occurs via two main pathways: progressive ductile deformation, which leads to loss of shape and function —-particularly at high temperatures and in flowable materials~\cite{zhang2021temperature, yao1998extensional}—- or the nucleation and propagation of localized cracks, which abruptly split the material~\cite{dowling_mechanical_2013,anderson_fracture_2017}. Cross-linked polymer networks, such as hydrogels and elastomers, cannot flow and thus fail exclusively by cracking.

Sequentially polymerized interpenetrated polymer networks, referred here as multiple polymer networks, including double- and triple-network hydrogels and elastomers, have emerged as a general and robust strategy to simultaneously enhance the stiffness and toughness of soft materials while retaining reversible elastic behavior at small strains. Since the seminal work by Gong and co-workers ~\cite{gong2003double, gong2010double}, double-network (DN) hydrogels have become prototypical examples of how soft polymer networks with remarkable mechanical properties can be designed by tuning network architecture, more specifically by combining a brittle, densely cross-linked network with a soft, extensible matrix, resulting in materials that are not only stretchable but also extremely tough. This architecture enables a remarkable combination of strength, extensibility, and resilience~\cite{xin2022review,zhao_multiscale_2014}. The prestretched sacrificial network fractures early, dissipating energy, while the matrix preserves structural integrity and enables delayed~\cite{gong2010double, ducrot2014toughening,matsuda_revisiting_2021,zheng2021chain,zhang_unique_2022} and more ductile~\cite{millereau2018mechanics,ju_role_2024} failure.

Inspired by these hydrogels, multiple-network elastomers have been developed to achieve enhanced performance in dry or swollen systems~\cite{creton2020multinet}. Toughness in these materials arises from controlled damage in the stiff network and stress transfer to a more extensible one. This strategy has produced elastic materials with fracture energies and failure strains far beyond those of their single-network counterparts~\cite{Slootman2022}.

Extensive experimental work confirms that bond scission in the sacrificial network plays a central role in the enhancement of the mechanical properties. Mechanophore-based imaging, fluorescence techniques, and postmortem analyses have shown that damage can extend hundreds of microns ahead of a crack tip, forming a broad damage zone ~\cite{Slootman2020, Slootman2022, Matsuda2020}. Yet, the microscopic mechanisms governing stress redistribution within and between networks in that dissipative zone—and how this leads eventually to damage localization—remains poorly understood.

Numerical simulations are a powerful tool to understand rupture in polymer systems~\cite{tauber2021sharing,fielding2025toughness,mugnai_interspecies_2025,masubuchi2025reviewmolecularsimulationsrupture}. Recent studies on multiple networks emphasize the importance of non-affine and heterogeneous stress redistribution. Tauber et al.~\cite{tauber2021sharing} showed, via coarse-grained molecular dynamics, that fracture in DN elastomers proceeds via a two-step process involving intra- and inter-network load transfer. Their results revealed that affine deformation models overestimate bond breakage, underscoring the need for spatially resolved models. Similar findings from experiments showed that load sharing through the matrix network delays failure and suppresses localization~\cite{Slootman2020,Matsuda2020,Slootman2022,ju_role_2024}.

Most prior studies, however, have examined either macroscopic mechanical behavior or simulations that lack spatial resolution of the fracture process. A notable exception is the mesoscale model by Walker and Fielding~\cite{fielding2025toughness}, which employed a 2D athermal representation of tightly coupled DN systems. Their work demonstrated that strong inter-network connectivity—although generally considered disadvantageous for network toughening~\cite{nakajima2009true}—can suppress stress concentrations and promote diffuse failure. While insightful, these findings are difficult to generalize to more complex, three-dimensional systems with loosely connected networks, where entropic contributions and structural heterogeneity dominate the early stages of deformation.

In this work, we address these limitations by combining spatially and temporally resolved light scattering measurements of microscopic dynamics~\cite{duri_resolving_2009} with three-dimensional coarse-grained molecular dynamics simulations in the rubbery regime to investigate bond scission and network rearrangements in single- and double-network elastomers under uni-axial loading. 
Experimentally, this approach allows us to directly visualize the mm scale spatially distributed microscopic activity and bond breaking prior to failure, showing that DN systems exhibit early-onset, delocalized dynamics before crack propagation, while in single-network (SN) systems macroscopic fracture is controlled by localized bond scission. At the nm scale, we simulate a volume element inside the damage zone undergoing uni-axial tension, demonstrating that for DN the bond scission remains spatially homogeneous up to a much larger stretch, while the SN localizes the damage at lower stretch. These results demonstrate that sacrificial bond breaking alone does not account for toughness; rather, toughness arises from spatio-temporal stress redistribution and enhanced microscopic dynamics that delay damage localization and crack growth.

\begin{figure}[]
\centering
\includegraphics[width=1.\linewidth]{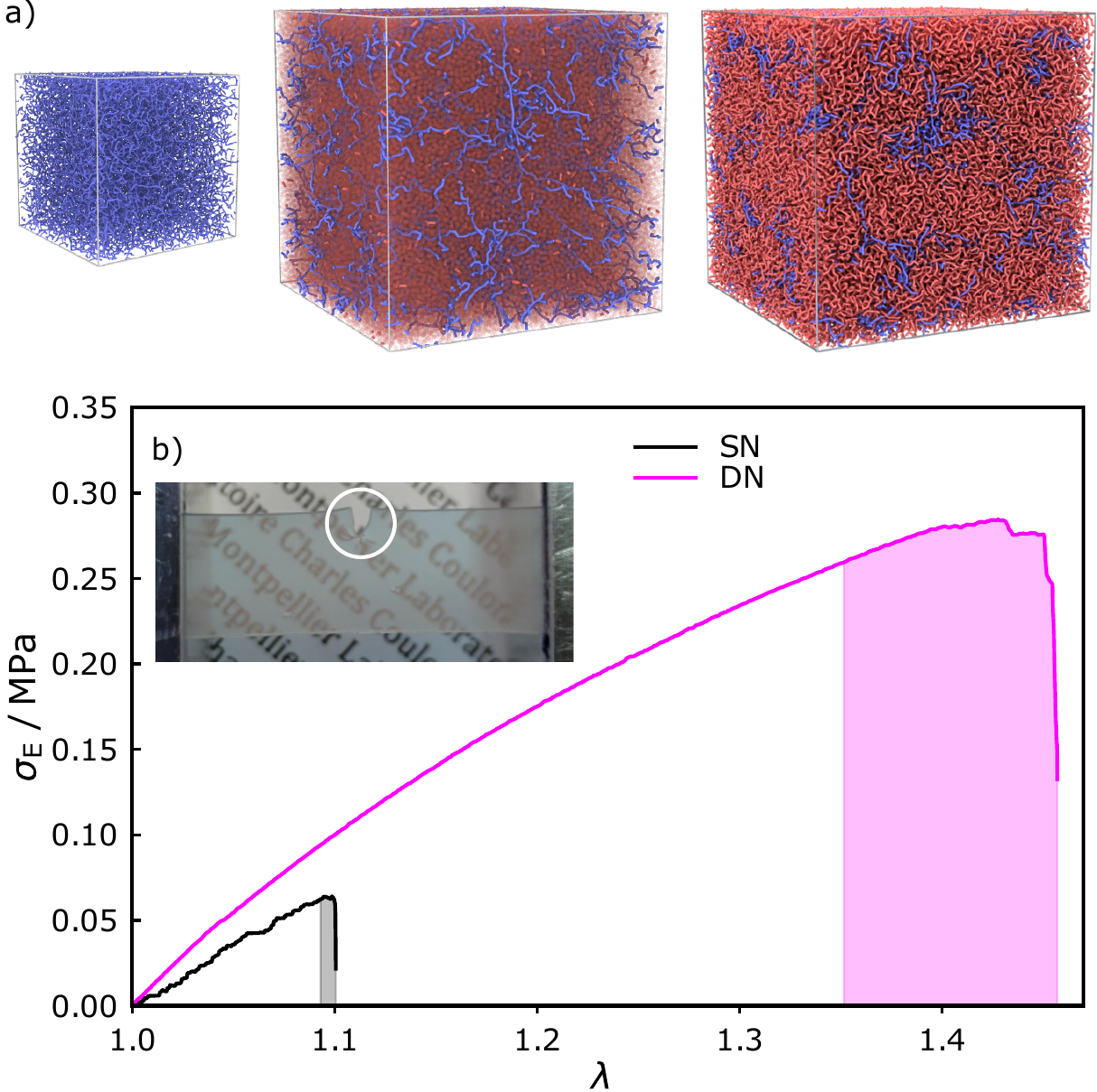}
\caption{\textbf{Architecture and mechanical properties of single and double polymer networks}. a) Architecture of the network model used in our simulations and obtained through a protocol analogous to the experimental one: a sacrificial network (blue, left snapshot) is first synthesized and then swollen by a linear factor $\lambda_0$ by adding monomers (red, middle snapshot), which are finally polymerized and cross-linked to form the matrix network (right snapshot). $\lambda_0 = 1.7, 2$ in our experiments and simulations, respectively. b) Main graph: experimental engineering stress $\sigma_\mathrm{E}$ \textit{vs} stretch $\lambda=L/L_0$ for single (SN) and double (DN) pre-notched networks, with $L$ and $L_0$ the sample length along the stretching direction during and at the beginning of the test, respectively. Shaded regions indicate where microscopic rearrangements increase significantly, as discussed in the text. 
Inset: image of a SN sample, the notch is circled. The sample size along the vertical direction is 10 mm.}
\label{fig:exp-stress-strain}
\end{figure}

\section*{Results}

We study experimentally and numerically single (SN) and double (DN) network elastomers. In experiments, poly (ethyl acrylate) networks are synthesized  as in~\cite{millereau2018mechanics}, see also \MM. Figure~\ref{fig:exp-stress-strain}a illustrates the main steps of the synthesis of a DN through snapshots of our numerical model of the networks (see \MM): a first, or sacrificial, network is osmotically swollen by a factor $\lambda_0$ by adding monomers ($\lambda_0=1.7$ in our experiments on DNs), which are in turn polymerized and cross-linked to form a second, or matrix, network, interpenetrating the first one.  Consistent with previous studies~\cite{millereau2018mechanics, ju_role_2024}, uni-axial tensile tests on pre-notched samples show that single and double networks have similar linear elasticity (Young's modulus = 0.84 and 1.27 MPa, respectively, see the \SuppInf), but DNs have significantly larger engineering stress, $\sigma_E$, 
and stretch, $\lambda = L/L_0$, at break, resulting in a remarkably enhanced toughness, see Fig.~\ref{fig:exp-stress-strain}b.

Simultaneously to the mechanical tests shown in Fig.~\ref{fig:exp-stress-strain}b, we perform Photon Correlation Imaging (PCI) measurements~\cite{duri_resolving_2009,orr_probing_2025a}, see Fig.~S2 of the \SuppInf, 
where images of laser light singly scattered by the networks during the traction test are recorded, allowing for sample visualization and quantification of the microscopic dynamics on multiple length scales from tens of nanometers to about 1 $\mu$m. 

\begin{figure}[]
\centering
\includegraphics[width=1.\linewidth]{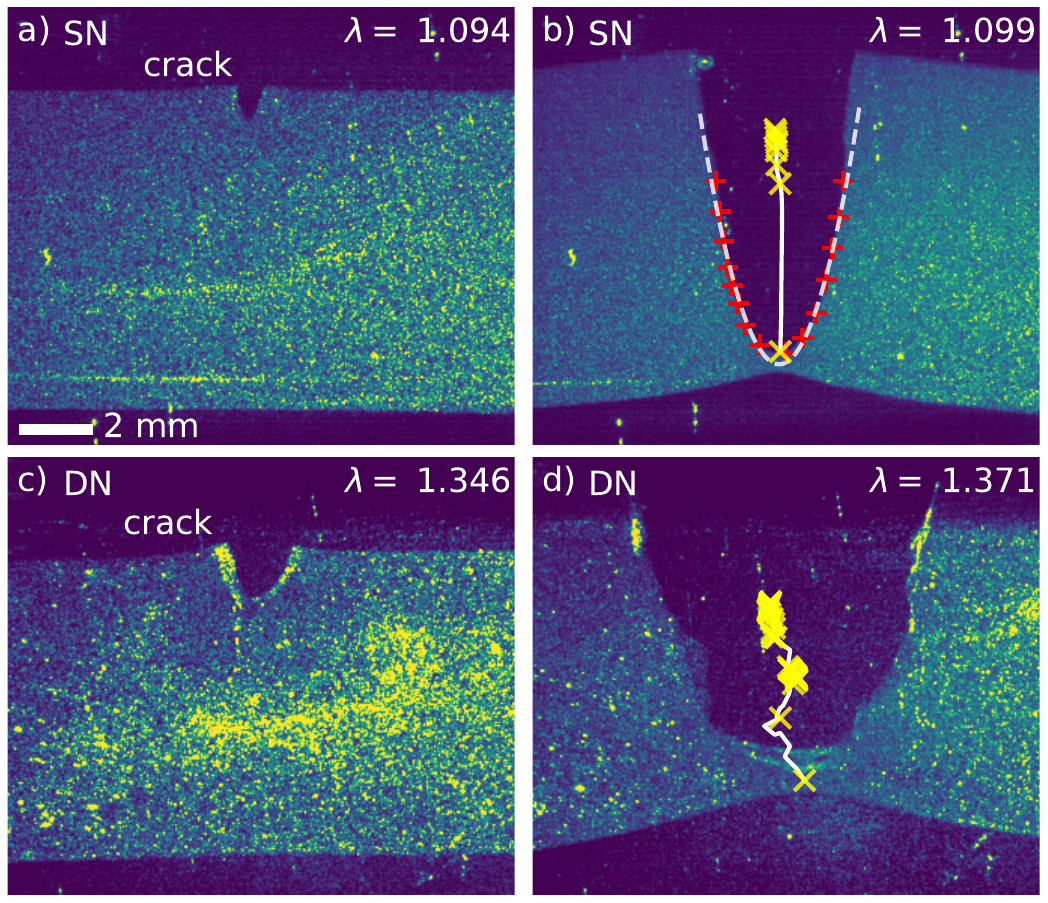}
\caption{\textbf{Notch propagation in single and double networks.} Colorized speckle images at selected $\lambda$ values, for a SN (a,b) and a DN (c,d), showing the crack shape and trajectory in a traction test. In b) and d), the white solid line is the crack tip trajectory, while yellow crosses show its position at intervals of one minute, starting from the position of the crack tip at the $\lambda$ value of the corresponding left panels. The dotted line in b) is a parabolic fit to the red pluses marking the crack shape. While the cracks are similar in single and double networks at the onset of propagation (a,c), they differ in shape and trajectory as they propagate through the sample.
}
\label{fig:exp-notch}
\end{figure}

Figure~\ref{fig:exp-notch} contrasts the shape and trajectory of the crack for a SN (a,b) and a DN (c,d) during a tensile test at fixed pulling speed on pre-notched samples. For the SN, the crack tip follows a straight trajectory, with progressively increasing speed. Furthermore, the crack shape is parabolic (Fig.~\ref{fig:exp-notch}b). These features are typical of crack propagation in a homogeneous elastic medium, where failure occurs at the crack tip due to stress concentration~\cite{hui2003crack,anderson_fracture_2017}. Although the crack shape in the DN is initially similar to that of the SN (Fig.~\ref{fig:exp-notch}a,c), it rapidly develops distinctive features: the shape of the crack becomes irregular and its trajectory meanders, 
alternating quasi-static periods with rapid displacements, see the yellow crosses in Fig.~\ref{fig:exp-notch}d. These features suggest that the crack progressively connects pre-damaged regions ahead of the crack tip that occur in randomly dispersed weak regions of the sample, hinting at more heterogeneous mechanical properties and distinct stress redistribution mechanisms in DNs as compared to SNs \cite{slootman:tel-02864025,ju_role_2024}.

To investigate failure at the microscopic level, we analyze the change in speckle intensity of the PCI images over a small stretch increment $\Delta \lambda$. Intensity fluctuations result from microscopic polymer chain displacements: We introduce a dimensionless dynamic activity $A
$, which quantifies microscopic displacements beyond those associated with the evolution of the macroscopic deformation field in the initial linear stretching regime, see \MM. We collect PCI images formed by light scattered in the forward (FS) and backward (BS) directions, thereby probing rearrangements on distinct length scales: $A \lesssim 0.8$, as typically observed in our experiments, corresponds to microscopic displacements $\lesssim \qty{1}{\micro\meter}$ and $\lesssim 70~\mathrm{nm}$ for FS and BS, respectively. As detailed in the~\MM, the activity $A$ depends on $\lambda$, the stretch increment $\Delta \lambda$ (or, equivalently, the time lag $\tau$ between PCI images), and location $\vec{r}$ in the sample. Activity maps are coarse-grained by averaging over regions of lateral size 0.8 mm by 0.8 mm and through the entire thickness of the sample. In all experiments, we fix $\tau = 8~\mathrm{s}$, corresponding to $\Delta \lambda \approx 8 \times 10^{-5}$.

\begin{figure}[h]
\centering
\includegraphics[width=1.\linewidth]{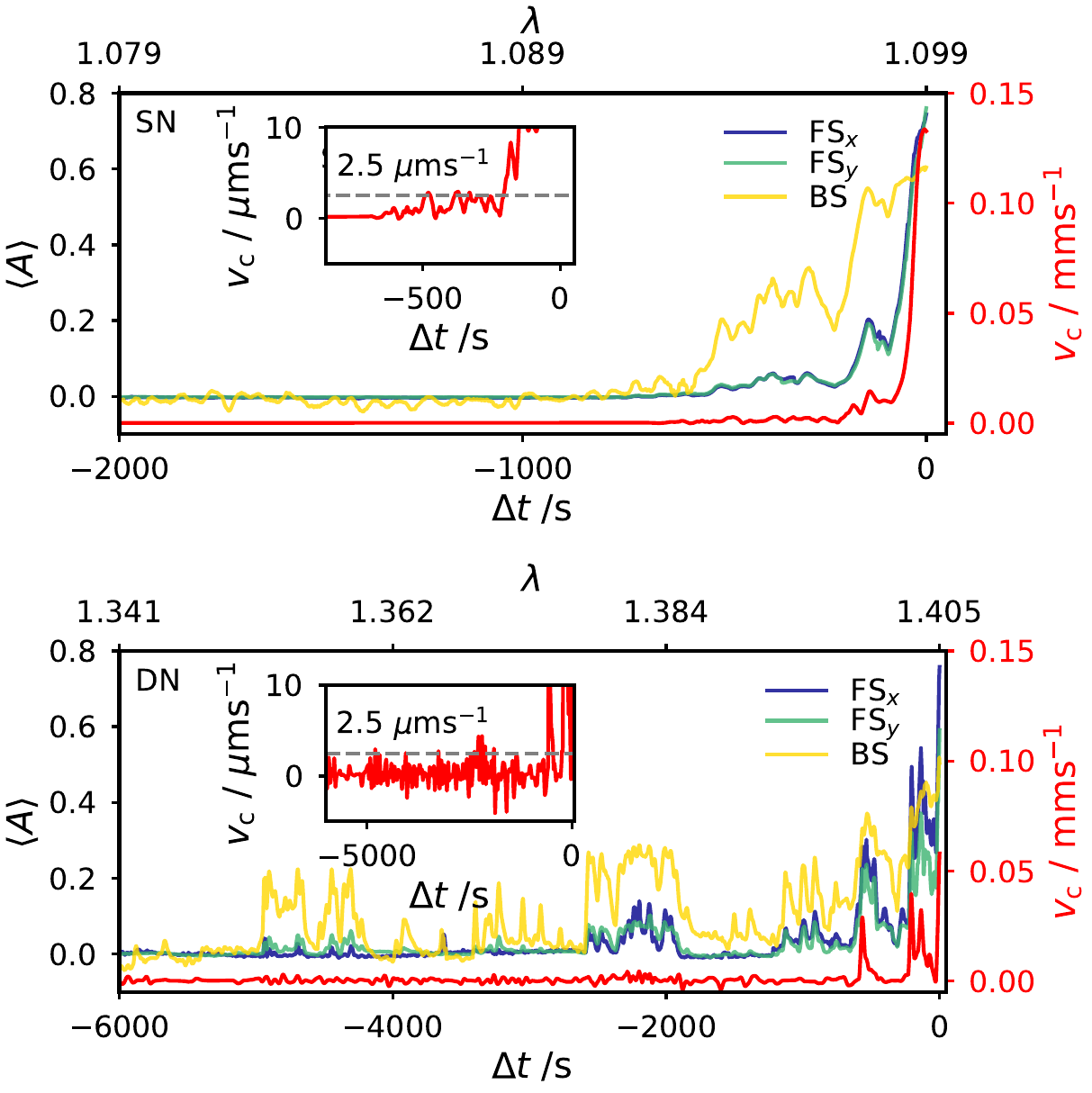}
\caption{\textbf{Crack speed and microscopic dynamics prior to rupture.} Left axis: time dependence of the microscopic rearrangements as quantified by the activity $A$ (see text for definition) averaged over the half-plane ahead of the crack tip, for scattering in the forward (FS$_x$, FSy) and backward (BS) directions, see \MM~and \SuppInf~for details. $\Delta t=0$ corresponds to macroscopic rupture and the top and bottom panels refer to representative single and double networks, respectively. The right axis and red line show $v_\mathrm{c}$, the propagation speed of the crack tip, with zooms on the behavior right before rupture in the insets. The dotted line in the insets is the threshold below which $v_\mathrm{c}$ is considered to be negligible for the analysis of Fig.~\ref{fig:exp-dyn-act-maps}. 
}
\label{fig:exp-dynact_vs_time}
\end{figure}

We start by focusing on $\left<A\right>$, the activity averaged over the sample volume ahead of the crack tip, shown in Fig.~\ref{fig:exp-dynact_vs_time} for a single and a double network (top and bottom panels, respectively), for BS and two orthogonal FS directions, probing microscopic motion along (FS$_x$) and perpendicular to (FSy) the pulling direction, see \MM. Data are plotted against time to macroscopic rupture (bottom axis), with the corresponding stretch shown on the 
top axis. For both the SN and the DN, $\left<A\right>$ is initially very close to zero, indicative of microscopic dynamics essentially identical to those in the mechanical linear regime. As $\lambda$ increases, $\left<A\right>$ rises above zero, signalling microscopic rearrangements due to local network damage. For both systems, the onset of microscopic rearrangements precedes macroscopic failure, similarly to what was reported recently for a variety of soft systems~\cite{cipelletti_microscopic_2020}, including colloidal or polymeric networks~\cite{leocmach_creep_2014,perge_time_2014,rogers_echoes_2014,colombo_stress_2014,landrum_delayed_2016,bouzid_elastically_2017,aime_microscopic_2018,kooij_laser_2018,filiberti_multiscale_2019,ju_real-time_2023}. Strikingly, however, the behavior of $\left<A\right>$ is very different for single and double networks. For the former, the onset of microscopic rearrangements occurs only a few hundreds of seconds before macroscopic rupture (see also the shaded region in Fig.~\ref{fig:exp-stress-strain}b ), with a nearly continuous increase of activity that accompanies the increase of the crack tip propagation speed $v_\mathrm{c}$ (red line and right axis). By contrast, for the DN, $\left<A\right>$ increases above zero as early as thousands of seconds before sample rupture, exhibiting intermittent bursts of activity, often not concomitant with crack propagation. For both the SN and the DN, the activity is higher in BS than in FS, due to the higher sensitivity of back-scattering to microscopic displacements (as discussed in detail in the~\MM). Inspection of the fracture of distinct samples of SN and DN suggests that these differences are more marked for the SNs, compare Figs. SI3 and SI4. It has been argued that in amorphous solids a weaker length-scale dependence of the dynamics stems from more numerous and widespread sources of local stress~\cite{martinelli_reaching_2023}. Thus, our PCI data hint to a larger process zone for the DNs than for the SNs. Interestingly, we find no significant differences for the two FS directions, indicating that microscopic rearrangements, while induced by uni-axial traction, are isotropic. 

\begin{figure}[h]
\centering
\includegraphics[width=1.\linewidth]{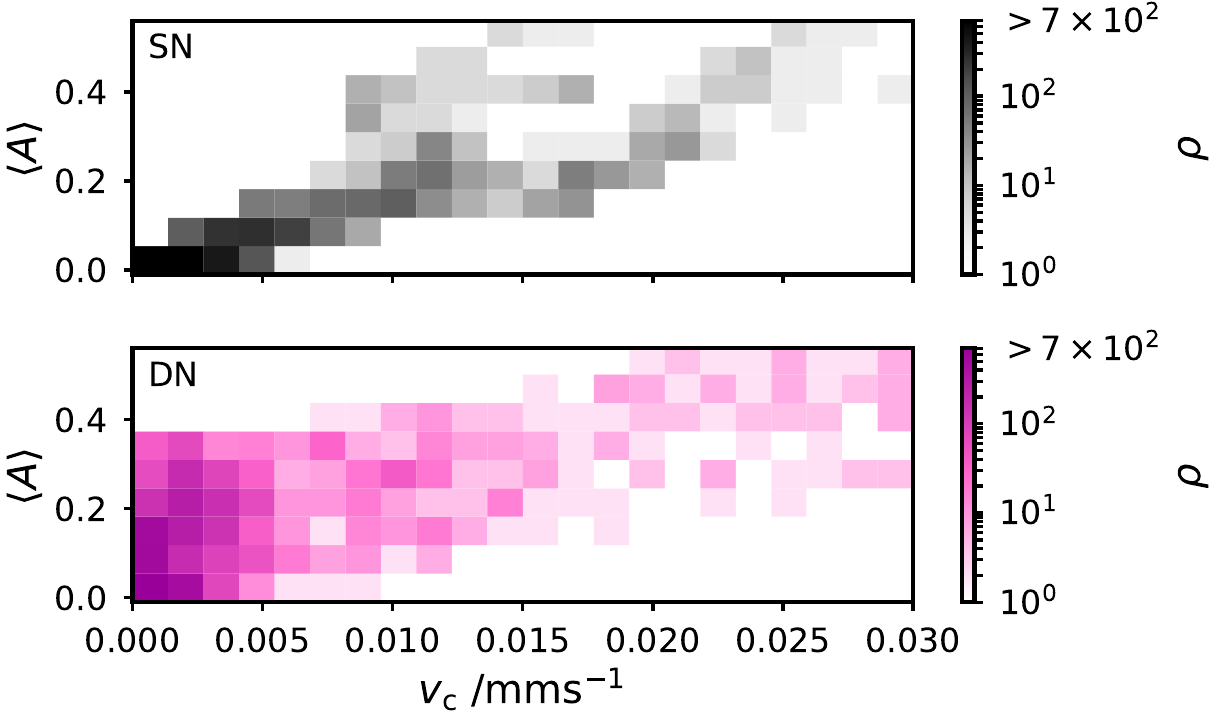}
\caption{\textbf{Correlation between crack propagation and microscopic dynamics depends on network architecture.} Color-coded number of occurrences of ($v_\mathrm{c}$, $\left<A\right>$) pairs obtained by binning measurements as those shown in Fig.~\ref{fig:exp-dynact_vs_time}, for SNs (top) and DNs (bottom), in the FS$_x$ geometry. Data obtained from tests on three SN and four DN samples in the 3000~s preceding macroscopic failure. Note the excess of large values of activity at vanishing $v_\mathrm{c}$ in DNs as compared to SNs. 
}
\label{fig:exp-dyn-act-vs-notch-speed}
\end{figure}

PCI is sensitive to any microscopic motion, including the relative displacement of network strands following a change in the macroscopic strain field upon crack propagation. To clarify the origin of the activity increase shown in Fig.~\ref{fig:exp-dynact_vs_time}, we inspect the histogram of the joint occurrence of pairs $(v_c,\langle A \rangle)$ approaching rupture. The results obtained by binning FS$_x$ data from all our experiments are shown in Fig.~\ref{fig:exp-dyn-act-vs-notch-speed} (see Fig. SI5 in the \SuppInf~for data for the two other scattering geometries). For the SN, the crack speed and activity are correlated, with no activity when there is no crack propagation. This suggests that in SNs the microscopic motion measured in our experiments is mostly due to the change of the strain field upon crack propagation, in analogy to recent findings for (single) networks of polydimethylsiloxane~\cite{ju_real-time_2023}. By contrast, for the DN a significant activity is observed at low crack propagation speeds and even without crack propagation, indicative of microscopic rearrangements not related to the change of the macroscopic strain field upon crack propagation. This difference is confirmed from measurements in the FS$_y$ and BS geometries, see Fig. SI5 in the \SuppInf.
These observations suggest that in DNs molecular damage blunts the crack, as opposed to making it propagate as in SNs.

\begin{figure}[h]
\centering
\includegraphics[width=1.\linewidth]{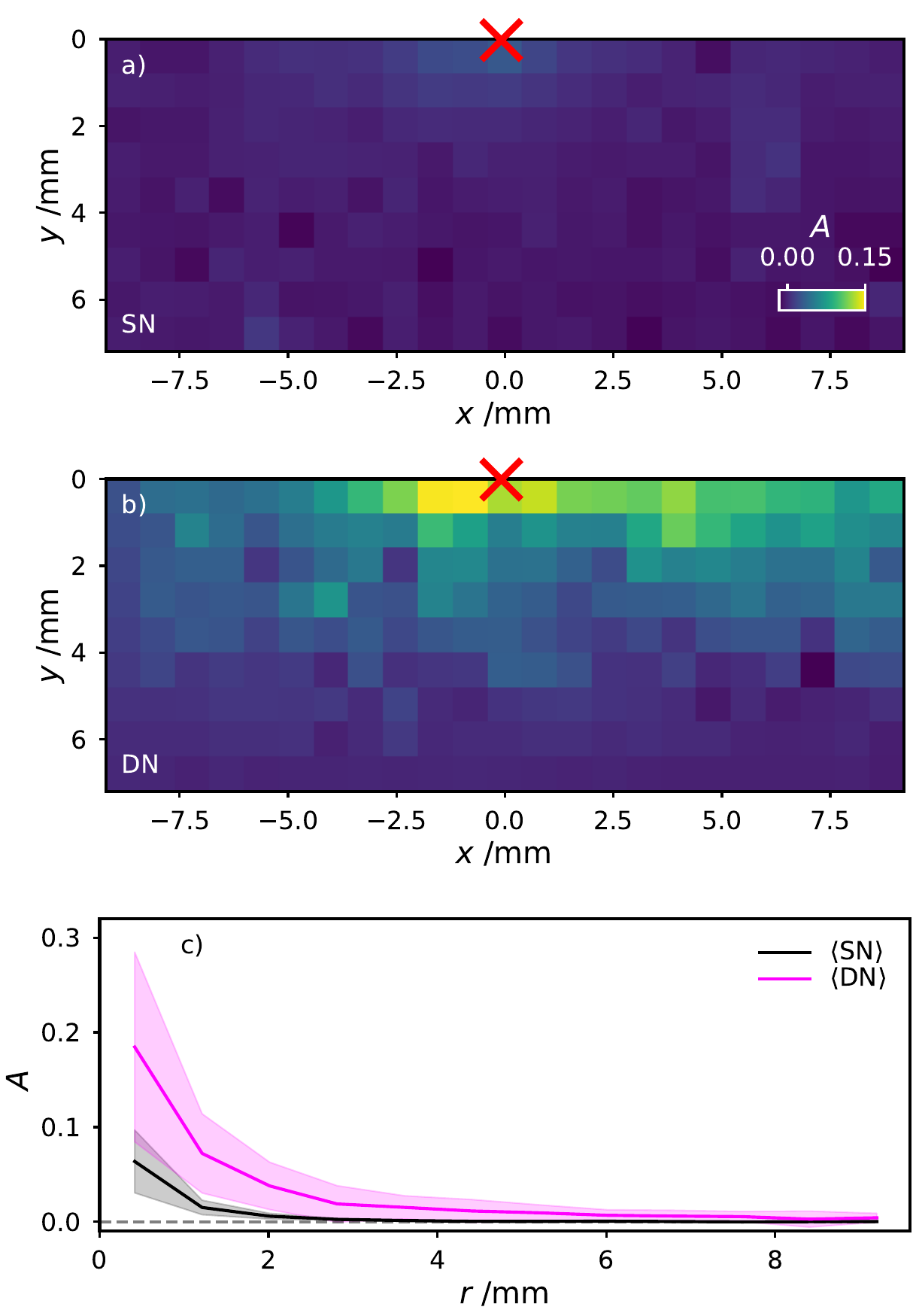}
\caption{\textbf{Spatial dependence of the activity in the half plane ahead the crack tip}. Spatial maps of $A$, for SNs (a) and DNs (b), in a moving reference frame where the crack tip corresponds to $(x,y) = (0,0)$ (red cross). Data are obtained by averaging FS$_x$ measurements for all samples, in time intervals with negligible notch propagation speed, $v_c \leq \qty{2.5}{\micro\meter\second^{-1}}$, and within 3000 s before macroscopic rupture. (c) Activity as a function of distance $r$ to the crack tip, obtained by azimuthally averaging the data of a) and b). The shaded areas show the standard deviation resulting from sample to sample variations.}
\label{fig:exp-dyn-act-maps}
\end{figure}

We examine the spatial structure of the rearrangement dynamics by building maps of $A$ in a moving reference frame where the crack tip is at the origin, averaging over multiple samples in the 3000 s preceding macroscopic fracture. As seen in Figs.~\ref{fig:exp-dyn-act-maps}a,b, we find a marked difference between SNs and DNs: for SNs, microscopic dynamics are overall less pronounced and more localized close to the crack tip, as compared to DNs. The dependence of $A$ on distance from the crack tip, obtained by azimuthally averaging the maps in a) and b), is shown in Fig.~\ref{fig:exp-dyn-act-maps}c. 
While for the SNs $A$ decays to zero essentially within the range of the spatial resolution of our maps, DNs exhibit non negligible microscopic dynamics up to a few millimeters from the crack tip. The results shown here refer to the FS$_x$ channel, which probes dynamics on a length scale $\sim \qty{1}{\micro\meter}$ in the direction parallel to pulling. Similar results are also obtained for the spatial dependence of FS$_y$ (dynamics perpendicular to the pulling direction, see Figs. S6a-c), while on the small length scale probed by the BS channel ($\sim70~$nm), the spatial maps of the dynamics for DNs are quite similar to those for SNs, with dynamics localized close to the crack tip, see Figs. S6d-f. The fact that for DNs microscopic rearrangement dynamics are seen over regions extending up to a few mm from the crack tip is surprising,  
in particular in view of experiments on polymer networks probing bond scission at the molecular level via fluorophores, suggesting little if any bond breaking at distances larger than about one hundred $\qty{}{\micro\meter}$ from the crack tip~\cite{ducrot2014toughening,Slootman2020,Slootman2022}. 
PCI measurements thus suggest that network rearrangements induced by bond breaking may extend over a much larger area than the directly damaged region.

To summarize our experimental findings, PCI measurements simultaneous to mechanical uni-axial traction tests unveil microscopic rearrangement dynamics that progressively build-up and dramatically accelerate hundreds to thousands of seconds before failure. These dynamics are strongly different for SNs and DNs, with respect to their temporal fluctuations, their correlation to crack propagation, and their spatial distribution. The emerging picture is that in DNs delocalized (both in time and space) plastic events protect the system from catastrophic failure associated with damage accumulation at the crack tip, as observed in SNs.

\begin{figure}[t]
\centering
\includegraphics[width=1.\linewidth]{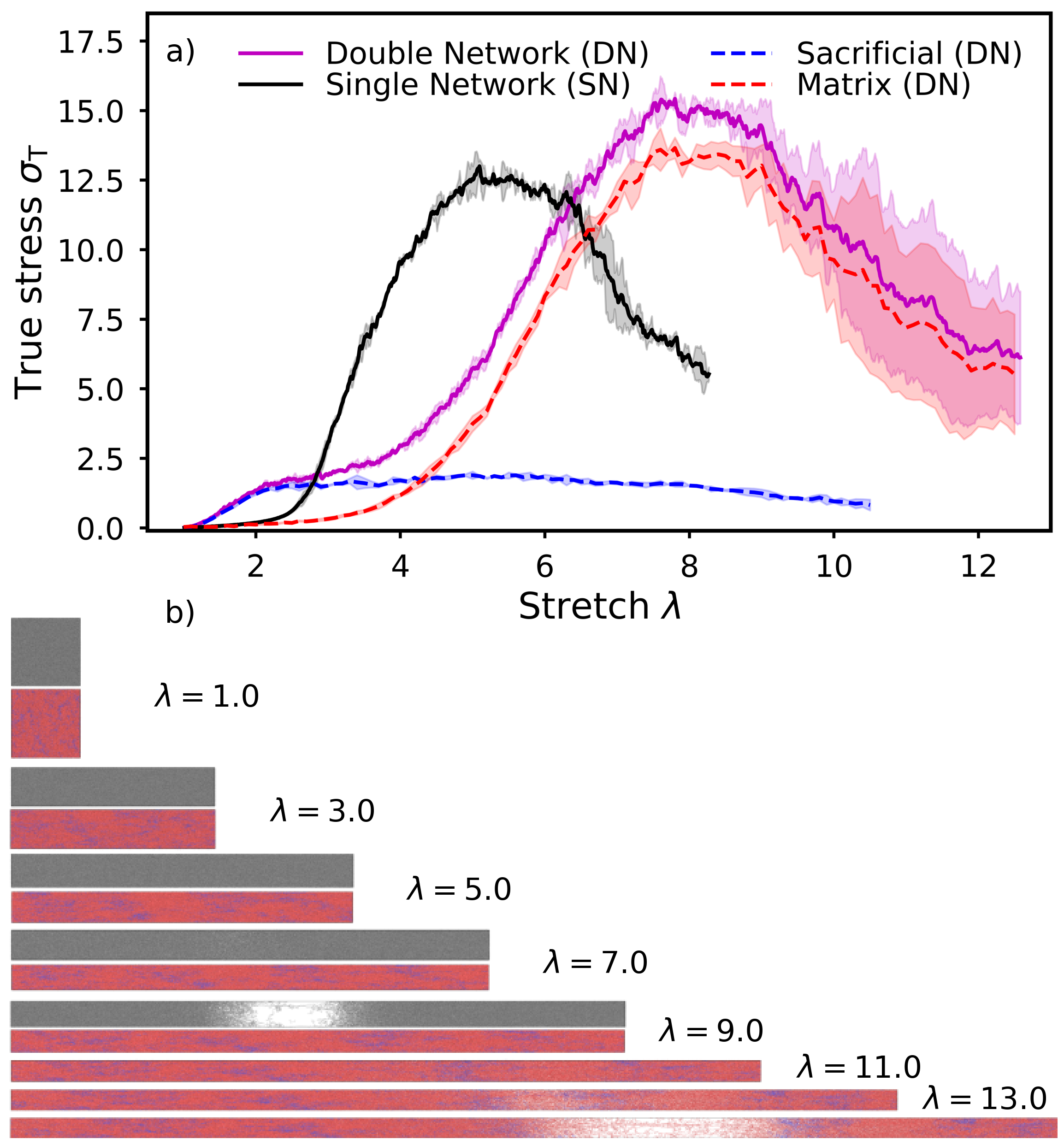} 
\caption{\textbf{Differences in mechanical behaviour and damage propagation between simulated single and double networks} (a) Stress $\sigma_\mathrm{T}=\sigma_{xx}-0.5(\sigma_{yy}+\sigma_{zz})$ as a function of stretch $\lambda$ for single (solid black line) and double (solid magenta line) networks. The stress contribution of the sacrificial and matrix sub-networks in the DN are shown with blue and red dashed lines, respectively. 
(b) Simulations snapshots projected onto the $(x,y)$ plane 
at various stretch values $\lambda$ for a SN (gray beads) and a DN (red and blue beads for the matrix and sacrificial networks, respectively).
}
\label{fig:sim-stress-strain-density-loc-1}
\end{figure}

To better understand at the microscopic scale the mechanisms imparting DNs their remarkable properties, we perform 3D molecular dynamics simulations, focusing on the structural rearrangements and stress redistribution of single and double networks submitted to uni-axial deformation. The simulations are coarse-grained at the polymer segment level rather than being fully atomistic, in order to study larger systems, see~\MM~for details. They capture the key differences in mechanical properties and damage propagation seen in the experiments: Fig.~\ref{fig:sim-stress-strain-density-loc-1}a shows that the maximum stress $\sigma_\mathrm{max}$ and the corresponding stretch $\lambda(\sigma_\mathrm{max}) = \lambda_\mathrm{peak}$ are larger in DNs relative to SNs, as seen in experiments (Fig.~\ref{fig:exp-stress-strain}b). Figure~\ref{fig:sim-stress-strain-density-loc-1}b shows snapshots of the simulated networks during stretching. Although macroscopic, pre-notched samples cannot be reproduced numerically, the simulated stretched volume can be seen as being representative of a volume element in the damaged zone preceding a crack~\cite{ju_role_2024}. The macroscopic fracture that eventually occurs in simulations, resulting from large, localized density fluctuations, can thus be pictured as the region where the damage zone fails, as discussed in two seminal theoretical papers on fracture of double networks~\cite{tanaka_local_2007,brown_model_2007}. Note that in the DN, significant damage occurs at larger $\lambda$ and is spread over a larger stretch interval as compared to the SN, mirroring the behavior of real samples, see the shaded regions in Fig.~\ref{fig:exp-stress-strain}b,  where a more stretched and hence wider damage zone results in a higher fracture energy~\cite{brown_model_2007}.

\begin{figure}[h]
\centering
\includegraphics[width=1.\linewidth]{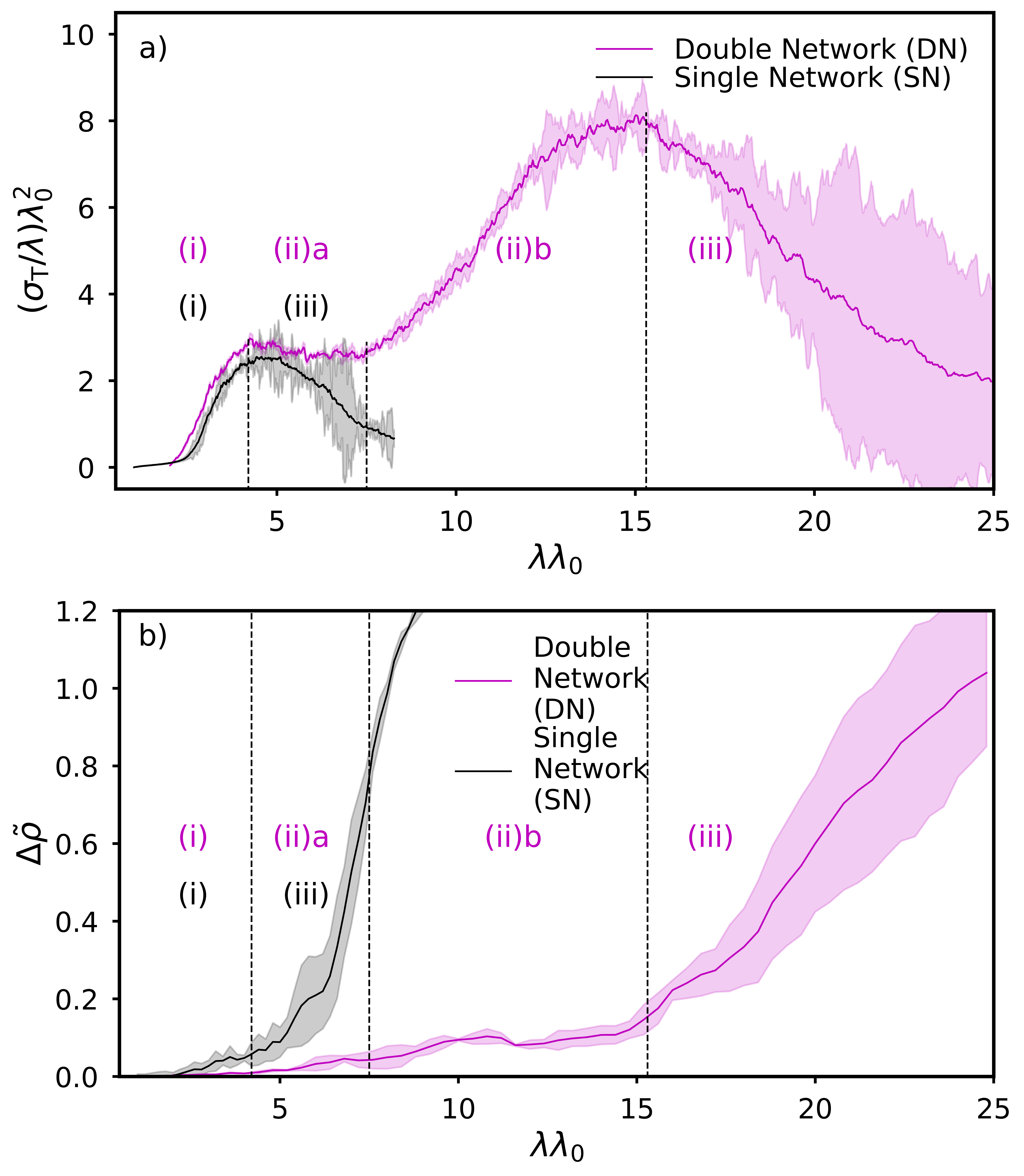} 
\caption{\textbf{Stress build-up and damage localization in simulated networks.} a): Rescaled stress-stretch curve (engineering stress $\sigma_\mathrm{T}/\lambda$ normalized by the areal density of sacrificial network strands $\lambda_0^{-2}$ \textit{vs} overall stretch of the sacrificial network, accounting for pre-stretch $\lambda_0$, with $\lambda_0 = 1, 2$ for the SN and DN, respectively.
b): Difference $\Delta \tilde{\rho}$ between the maximum and the minimum of density along the stretching direction (averaged over the orthogonal direction $y$ and $z$) \textit{vs} rescaled stretch $\lambda\lambda_0$.
}
\label{fig:sim-stress-strain-density-loc-2}
\end{figure}

In order to highlight similarities and differences between single and double networks, we plot in Fig.~\ref{fig:sim-stress-strain-density-loc-2}a the same data as in Fig.~\ref{fig:sim-stress-strain-density-loc-1}a, but using rescaled variables that account for $\lambda_0$, the swelling-induced prestretching of the sacrificial network embedded in the DN~\cite{millereau2018mechanics}, where $\lambda_0=2$ for the simulated DNs. In this representation, the rescaled strain-stretch curves of both systems initially nearly superimpose, indicating that at small stretch the elasticity of DNs is essentially ruled by the response of the sacrificial network, as confirmed by examining separately the various contributions to the stress in the DN, see the red and blue dashed lines in  Fig.~\ref{fig:sim-stress-strain-density-loc-1}a. This representation also allows for clearly identifying distinct regimes during the test. For the SN, an initial growth of the stress (regime (i)) is followed by a drop of $\sigma_\mathrm{T}$ for $\lambda\lambda_0 \gtrsim 5$ (regime (iii)), eventually resulting in breakage. The DN exhibits the same regimes, but separated by two intermediate regimes, (ii)-a and (ii)-b, where $\sigma_\mathrm{T}$ first plateaus and then increases again.

The various regimes highlighted in Fig.~\ref{fig:sim-stress-strain-density-loc-2}a correspond to specific changes in the structure of the networks, revealed in Fig.~\ref{fig:sim-stress-strain-density-loc-2}b, where we report the density fluctuations during stretching, quantified by $\Delta \tilde{\rho}$, the relative amplitude of the difference between the densest and sparsest network regions, see \MM~and Fig. S7 in the \SuppInf.
For both systems, regime (i) corresponds to a very mild growth of density fluctuations, due to isolated, rare bond breaking events. In this regime, bond rupture occurs randomly in space and is governed by the properties of the initial network configuration such as, e.g., the initial strand stretch \cite{tauber2021sharing}.
In regime (iii), density fluctuations grow rapidly in both systems. Note that in the simulations, where the system is not pre-notched, the growth of large density fluctuations leading to the formation of a macroscopic crack is akin to the failure in the damage zone at the tip of the crack in the experimental crack propagation regime.
The intermediate regimes (ii)-a and (ii)-b, seen in DNs but not in SNs, correspond to a mild growth of $\Delta \tilde{\rho}$ due to small clusters of breaking events and to the progressive damage of the matrix, respectively (see Fig S7 and S8 in \SuppInf). Interestingly, in regimes (ii)-a and (ii)-b the sacrificial network exhibits ductile behavior due to delocalized damage, as shown by the stress plateau over an extended range of stretch, dotted blue line in Fig.~\ref{fig:sim-stress-strain-density-loc-1}a. Thus, in DNs the matrix network protects the sacrificial network from \textit{brittle} failure, a first important observation to explain the origin of the remarkable mechanical properties of double networks.  

\begin{figure}[H]
\centering
\includegraphics[width=1\linewidth]{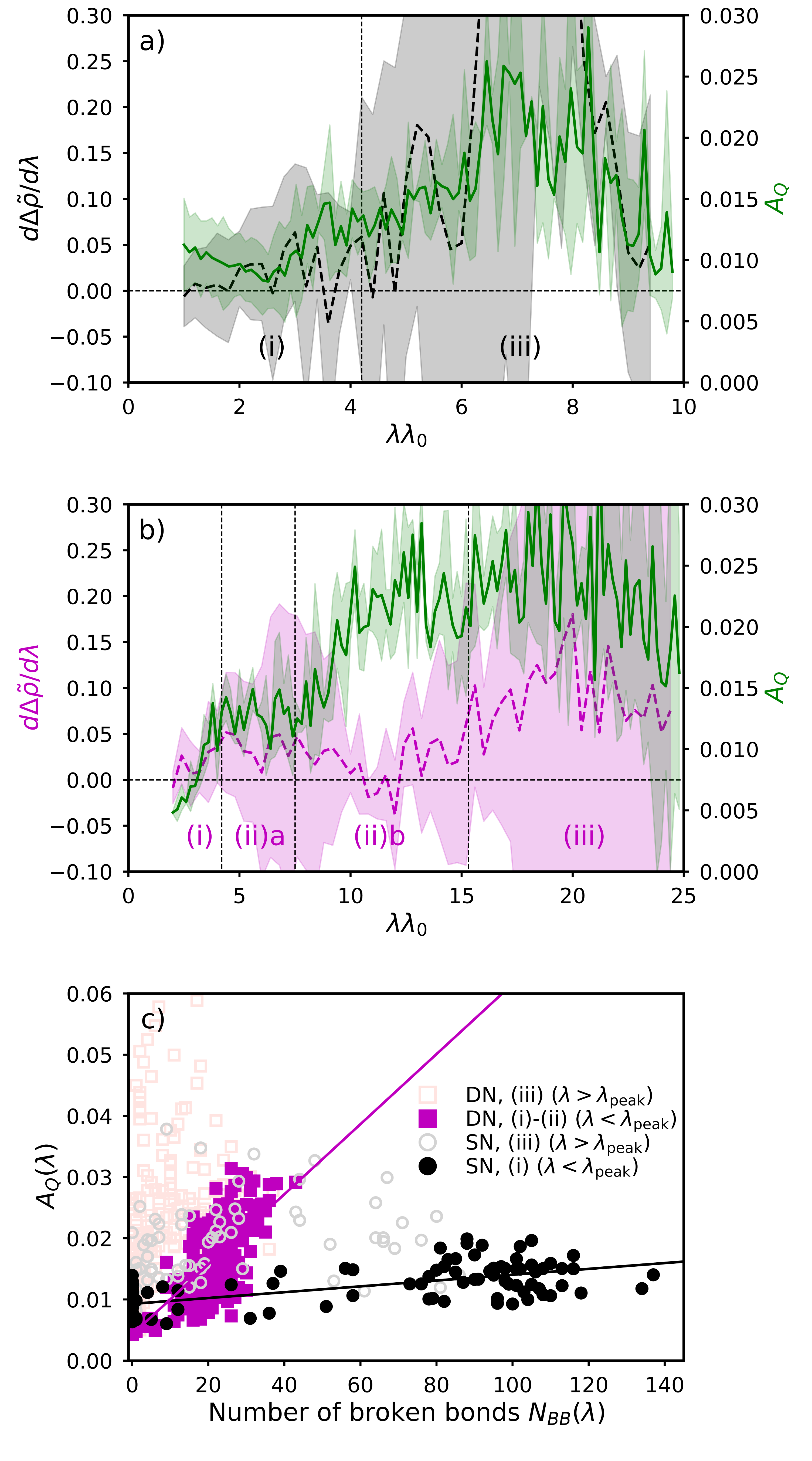} 
\caption{\textbf{Correlation between the microscopic dynamics and density localization and bond breaking in simulated single and double networks.} 
a),b): Derivative of density localization (dashed line and left axis) and dynamic activity
$A_Q$ (green line and right axis) \textit{vs} $\lambda \lambda_0$ for single, a), and double, b), networks. The shaded areas indicate standard deviation over two SN and three DN realizations.
c) Dynamic activity as a function of number of broken bonds, both calculated over a stretch increment $\Delta \lambda = 0.40$, for single (black circles) and double (magenta squares) networks, before (filled symbols) and after (empty symbols) the stress peak, respectively. 
The lines are linear fits to the data in the $\lambda < \lambda_\mathrm{peak}$ regime, $A_Q = s N_\mathrm{BB} + A^0$, with $s_\mathrm{SN} = 4.5 \times 10^{-5}$, $A_\mathrm{SN}^0 = 9.3 \times 10^{-3} $ and $s_\mathrm{DN} = 5.7 \times 10^{-4}$, $A_\mathrm{DN}^0 = 4.3 \times 10^{-3} $ for SNs and DNs, respectively. 
}
\label{fig:sim-dynamics-BB}
\end{figure}

In experiments, microscopic rearrangements with no crack propagation where seen for DNs, a feature absent in SNs (Fig.~\ref{fig:exp-dynact_vs_time}). This is confirmed in our simulations by direct measurements of damage propagation and microscopic motion. 
Microscopic dynamics are quantified by a dynamic activity $A_Q = 1-Q$ (green line and right axis in Figs.~\ref{fig:sim-dynamics-BB}a,b), akin to the experimental space-averaged dynamic activity $\left< A \right>$ discussed in relation to Fig.~\ref{fig:exp-dynact_vs_time}. $A_Q$ quantifies the loss of overlap $Q$ between successive sample configurations during stretching, spaced by a stretch $\Delta \lambda = 0.4$, chosen to filter out small-amplitude displacements associated to thermal fluctuations, see \MM. Damage propagation is quantified by $\mathrm{d}\Delta \tilde{\rho} / \mathrm{d}\lambda $, the rate at which density fluctuations grow upon stretching. Figures~\ref{fig:sim-dynamics-BB}a,b contrast the behavior of single and double networks. In the former, microscopic motion is strongly correlated to $\mathrm{d}\Delta \tilde{\rho} / \mathrm{d}\lambda$ (black dashed line and green solid line), throughout all regimes. By contrast, in the latter we observe significant microscopic dynamics with no corresponding damage propagation in regime (ii)-b of the DNs (left axis and dashed pink line). 

Simulations allow for establishing a direct link between microscopic dynamics and bond breaking, a feature that could not be directly measured in our experiments. Figure ~\ref{fig:sim-dynamics-BB}c shows the relationship between microscopic dynamics and the number $N_{\mathrm{BB}}$ of broken bonds, both calculated over small stretch increments $\Delta \lambda = 0.40$. 
Two different regimes are seen. For the solid symbols corresponding to strains smaller than $\lambda_\mathrm{peak}$ (defined in Fig.~\ref{fig:sim-stress-strain-density-loc-1}a), there is a clear linear correlation between $A_Q$ and $N_{BB}$: broken bonds are thus directly responsible for the observed dynamics. For $\lambda > \lambda_\mathrm{peak}$, by contrast, data are quite scattered (open symbols), with an overall trend of enhanced dynamics (larger $A_Q$) due to large local displacements associated with crack growth. Remarkably, we find that dynamics are overall enhanced in DNs: in the $\lambda < \lambda_\mathrm{peak}$ regime, bond breaking in DNs results in dynamics about five times larger than for SNs. For $\lambda > \lambda_\mathrm{peak}$, the increase is threefold, on average. These results are consistent with the experiments, where DNs overall exhibited enhanced dynamics as compared to SNs. Note that chain motion in the crowded environment of the networks entails chain friction, and hence viscous dissipation. Thus, the enhanced chain dynamics in DNs as compared to SNs indicates that energy dissipation is more important in DNs, a second key finding explaining the overall difference in the macroscopic mechanical response between single and double networks. 

\begin{figure}[H]
\centering
\includegraphics[width=1.0\linewidth]{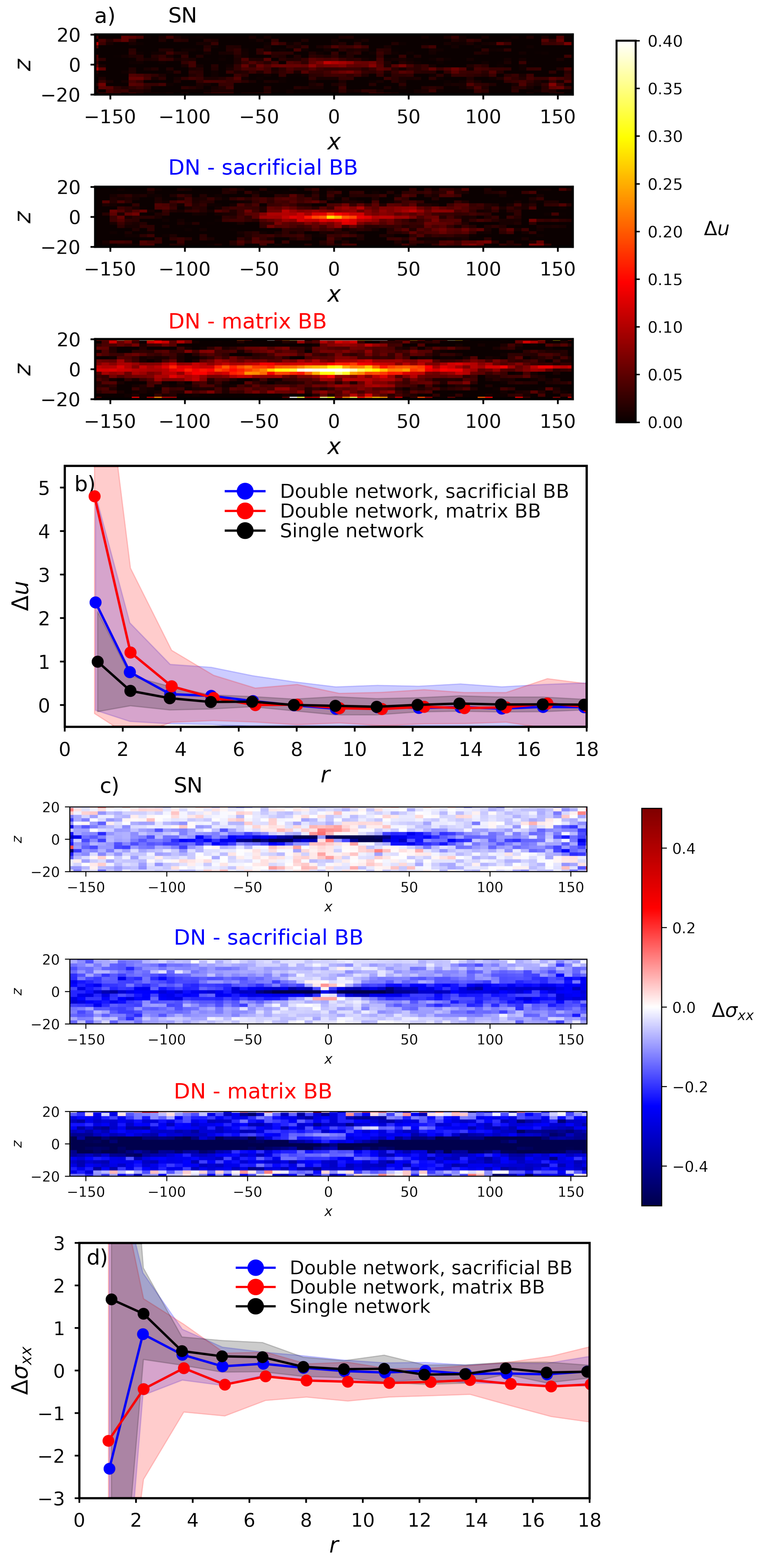} 
\caption{\textbf{Response to bond breaking: displacement field and stress relaxation.} 
a) Displacement magnitude, $\Delta u$, as a function of the $x$ and $z$ coordinates, in response to a single bond breaking event located in $(0,0,0)$ in the SN (top), and in the DN, for bond breaking occurring either in the sacrificial network (middle) or in the matrix (bottom). Data averaged over 40-80 events in each panel.
b): Displacement magnitude \textit{vs} distance $r$ to a bond breaking event, measured in the plane $x=0$ perpendicular to the stretching direction. c): Maps of the average change in the $xx$ component of stress following a single bond breaking event, see labels for the various networks. d) Average change in stress $\sigma_{xx}$ as a function of $r$, defined as in b). In b) and d), the shaded areas indicate the standard deviation over 40-80 bond breaking events.}
\label{fig:sim-stress-dyn-upon-BB}
\end{figure}


The enhanced dynamics seen in simulations of DNs stems from rearrangements involving displacements of larger magnitude and occurring over a larger region as compared to SNs. This is shown in Figs.~\ref{fig:sim-stress-dyn-upon-BB}a,b, which display the spatial dependence of the dynamics upon a bond breaking event, taking the origin as the event location (details in the \MM). Maps are obtained by averaging over 40-80 bond breaking events occurring in all regimes\footnote{For the SNs, the vast majority of bond breaking events occurs in regime (iii). For the DNs, a large majority of the bond breaking events in the sacrificial network occurs in regimes (i), (ii)-a and (ii)-b, while for the matrix network most events occur in regimes (ii)-b and (iii)}. Overall, the displacement field $\Delta u$ generated by a bond breaking event is larger and longer-ranged in the DNs as compared to the SNs, with remarkable differences between the sacrificial and matrix networks within a SN. Indeed, a bond breaking event in the matrix generates more displacement than one occurring in the sacrificial network (compare the bottom panel of Fig.~\ref{fig:sim-stress-dyn-upon-BB}) to the middle one), as also seen in Fig.~\ref{fig:sim-stress-dyn-upon-BB}b, which displays $\Delta u$ as a function of distance $r$ to the broken bond, in the $x=0$ plane perpendicular to stretching (details in the ~\MM). This may be due to the fact that the cross-link concentration in the matrix network is lower than in the sacrificial one (see the ~\MM), and hence the contour length of the chain segments in between cross-links is larger in the former than in the latter, allowing for exploring a larger number of configurations in response to a breaking event. This highlights the key role of the matrix network in dissipating energy following a bond breaking event, a new result generally overlooked in qualitative interpretations of the toughening effect.
 
The enhanced dynamics of the DNs result in a more efficient redistribution of the stress following a bond breaking event, as depicted in Figs.~\ref{fig:sim-stress-dyn-upon-BB}c,d, where we show $\Delta \sigma_{xx}$, the $xx$ component of the change in the stress tensor in response to a single bond breaking event. Here, we consider the stress variation due to the the change of the bond interaction potential, since this is the dominant contribution, see Fig. S11 in the \SuppInf. In Figure~\ref{fig:sim-stress-dyn-upon-BB}c, positive $\Delta  \sigma_{xx}$ (red shades) corresponds to bond loading and hence enhanced breaking probability, whereas negative $\Delta \sigma_{xx}$ (blue shades) indicates reduced bond stress, entailing a lower breaking probability. 
In the SNs, bond breaking leads to stress redistribution with a distinctive dipolar appearance, top panel of Fig.~\ref{fig:sim-stress-dyn-upon-BB}c, resulting in regions with significant increase of the load, as generally observed in elastic media~\cite{creton2016fracture}, which may cause additional bond breaking in an avalanche-like manner. A similar dipolar structure is also observed in the $\Delta \sigma_{xx}$ field when breaking occurs in the sacrificial network in the DN (mid panel of Fig.~\ref{fig:sim-stress-dyn-upon-BB}c), but stress release is enhanced  and spread over a larger region as compared to the SN.
Remarkably, the stress drop following a rupture event in the matrix network (bottom panel of Fig.~\ref{fig:sim-stress-dyn-upon-BB}c) is much larger than in the sacrificial network or the single network as expected from the larger cross-link concentration \cite{lake1967proc}. These differences are summarized in Fig.~\ref{fig:sim-stress-dyn-upon-BB}d, which displays $\Delta \sigma_{xx}(r)$. The stress change along the radial ($r = \sqrt{y^2+z^2}$) direction is positive and decays slowly with $r$ for the SN, indicating an overall increase of the stress in the direction perpendicular to stretching following an event, while for the DNs it is close to zero or even negative at small $r$, confirming efficient stress release and redistribution. Ultimately, the differences in stress redistribution highlighted in Fig.~\ref{fig:sim-stress-dyn-upon-BB} are responsible of the synergistic effects between the sacrificial and matrix networks leading to an efficient microscopic energy dissipation mechanism that helps sustaining much larger stresses and delays the onset of rupture.

\begin{figure}[H]
\centering
\includegraphics[width=1.\linewidth]{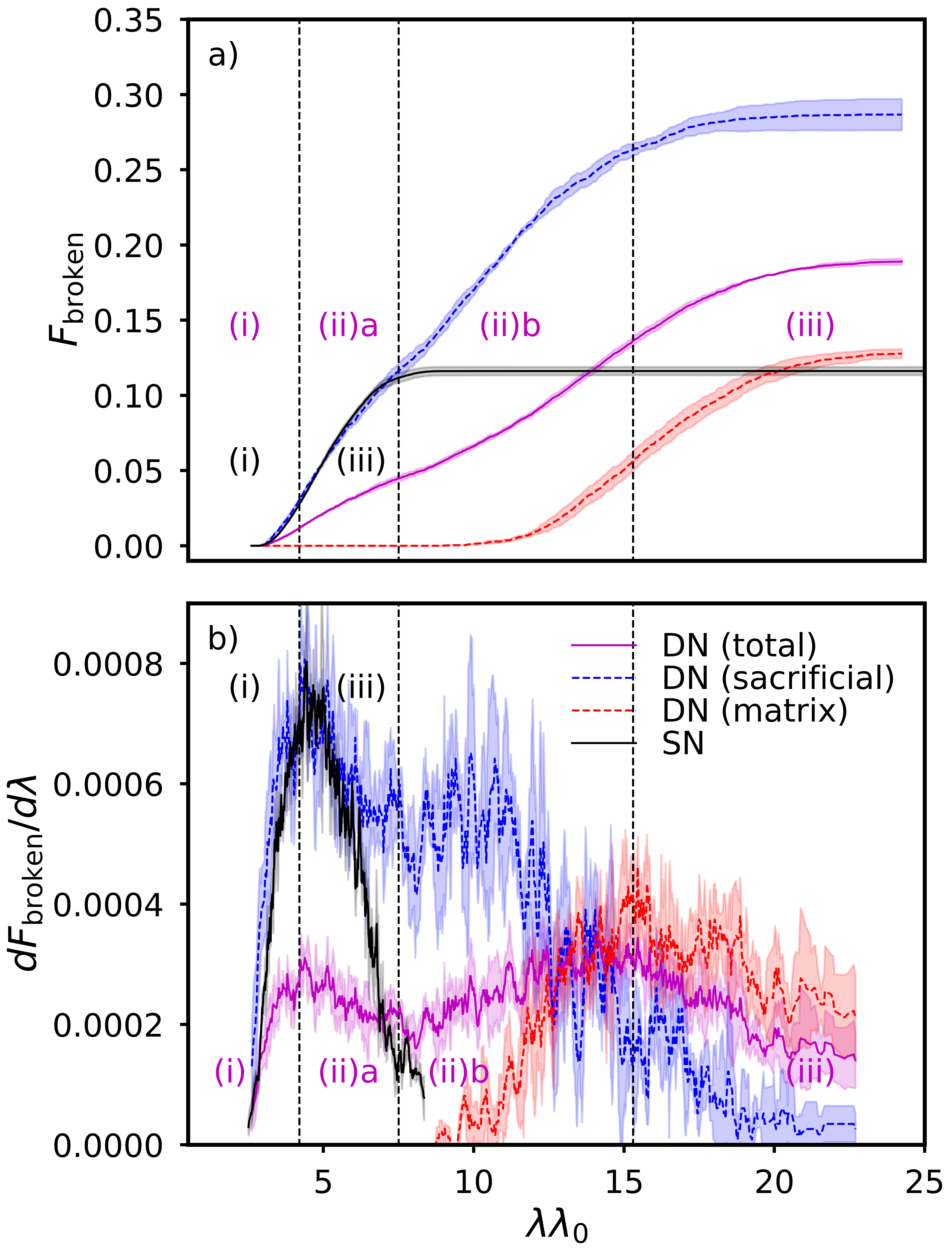} 
\caption{\textbf{Double networks are resilient to bond breaking.} a) Fraction of broken strands in the single (black solid line) and double (magenta solid line) networks \textit{vs} rescaled stretch $\lambda\lambda_0$. The contributions of the sacrificial (blue) and matrix (red) networks are shown as dashed lines. The fraction of broken strands is calculated for each network relative to its own total number of strands. b) Rupture rate as a function of rescaled stretch $\lambda\lambda_0$ (derivative of the data of panel a)). In a) and b), shaded areas indicate the standard deviation over different realizations for the DNs.
}
\label{fig:sim-BBdyn-TBC}
\end{figure}

These mechanisms endow the DNs with a higher resilience to bond breaking: as shown in Fig.~\ref{fig:sim-BBdyn-TBC}a, a larger fraction of the network bonds have to be broken for a DN to fail, as compared to SNs. The stretch evolution of $F_\mathrm{broken}$, the fraction of broken strands with respect to all strands in a given network, follows distinct behaviors in the various regimes defined in Fig.~\ref{fig:sim-stress-strain-density-loc-2}. 
For the SNs (black line), the rate of events grows steeply in regime (i) and then decreases rapidly in regime (iii), since a few localized events occurring around a zone where several bonds have already been broken are sufficient to lead to crack amplification and sample failure. As a consequence, $F_\mathrm{broken}$ grows rapidly and plateaus at a relatively low value, $F_\mathrm{broken} \approx 12\%$, at failure. For the DNs (magenta line), after an initial growth in regime (i), the rate of breakage remains moderate and nearly constant over an extended stretch range, $\lambda\lambda_0 \le 17$, showing that not only is damage spread over a wide zone, as seen in Fig.~\ref{fig:sim-stress-strain-density-loc-1}b, it also occurs more progressively in time as compared to SNs .
The dashed blue and red curves in Figs.~\ref{fig:sim-BBdyn-TBC}a show the contributions of the sacrificial and matrix networks, respectively. In regime (i), essentially all broken strands belong to the sacrificial network, whose evolution follows that of a SN: this is consistent with the observation that in regime (i) the matrix network is almost unloaded, Fig.~\ref{fig:sim-stress-strain-density-loc-1}a. This is seen also by inspecting the rate at which bonds are broken, shown in Fig.~\ref{fig:sim-BBdyn-TBC}b, which further reveals that in regime (ii)-a and in the initial phases of (ii)-b, there still are virtually no events in the matrix network, which protects the sample from macroscopic failure via the stress redistribution mechanisms discussed in Fig.~\ref{fig:sim-stress-dyn-upon-BB}. Interestingly, the bond breaking rate in the sacrificial networks exhibits a second maximum in regime (ii)-b, indicating that stress coupling between the two networks enhances sacrificial bond breaking. Crucially, this allows $F_\mathrm{broken}$ of the sacrificial network to keep growing, well above the level attained in SNs at failure. In the second part of regime (ii)-b, the rate of events in the sacrificial network declines, signalling incipient failure. At the same time, the matrix network takes over, as shown by the rapid increase of breaking events therein. These events occur where the sacrificial network has already been damaged, see Fig. S8 in the~\SuppInf, most likely because stress accumulates in the voids formed in the sacrificial network. Remarkably, the decrease in the rate of bond breaking in the sacrificial network is nearly balanced by the growth of bond breaking in the matrix network, resulting in the nearly constant breaking rates observed in the DN as a whole. Finally, $F_\mathrm{broken}$ levels off in regime (iii), as for the SN. 

\subsection*{Discussion and Conclusions}

Previous works investigating the origin of the outstanding toughness of multiple polymer networks pointed to the key role of energy dissipation through the enhanced scission of bonds, in particular in the sacrificial network~\cite{ducrot2014toughening,ju_role_2024,tauber2021sharing,fielding2025toughness}, and the subsequent stress redistribution. They also hinted at damage delocalization as a possible mechanism protecting multiple networks. Damage delocalization was reported in 2D simulations of a mesoscopic network model~\cite{tauber2021sharing,fielding2025toughness} and was included in another mesoscopic model proposed to account for the evolution of the strain field measured in uni-axial extension tests on multiple elastomer networks~\cite{ju_role_2024}. However, confocal imaging of bond scission via mechanophore cross-linkers suggested that in double networks similar to those studied here, the damage zone extends only over limited distances from the crack, of the order of $100~\mu$m~\cite{ducrot2014toughening,Slootman2020,Slootman2022}, leaving open the question of the microscopic mechanisms allowing for the effective energy dissipation in multiple elastomer networks.

Our experiments coupling the quantification of microscopic dynamics (PCI) and mechanical tests, unveil, for DNs, enhanced microscopic dynamics over millimeter-sized regions, well ahead of macroscopic failure. We propose that these dynamics, uncorrelated from crack propagation, are a key mechanism for energy dissipation. Dynamics extend as far as millimeters away from the crack tip, where bond scission is presumably too rare and evenly distributed to be directly measured by mechanophore imaging, explaining a lack of signal in previous works~\cite{ducrot2014toughening,Slootman2020,Slootman2022}. This should be contrasted with SNs, where microscopic activity is localized closer to the crack and is correlated with its propagation. Interestingly, we find that the difference between DNs and SNs in the range of the microscopic activity and in its correlation to crack propagation is more marked in forward scattering than in backward scattering, compare Figs.~\ref{fig:exp-dyn-act-vs-notch-speed} and~\ref{fig:exp-dyn-act-maps} to Figs. S5 and S6, respectively. This suggests that in DNs dissipation is enhanced mainly through the relaxation of density fluctuations on large length scales, of the order of 1 µm, while dynamics on smaller length scales are less differentiated in single and double networks. This highlights the importance of a multiscale approach, which goes well beyond the molecular scale of bond scission and even the ten-nanometer scale probed in backscattering or in previous works based on multiple scattering experiments~\cite{kooij_laser_2018,ju_real-time_2023}.


Our simulations support these qualitative differences and help to rationalize the experimental results. We have shown that density fluctuations in DNs are distributed over a significantly larger region than in SNs. This confirms our experimental finding that bond scission may affect larger sample volumes than previously inferred from mechanophore measurements~\cite{ducrot2014toughening,Slootman2020,Slootman2022}. This discrepancy may be due to the fact that scission events spread homogeneously over length scales larger than the confocal microscopy resolution, which would result in a weak and difficult-to-detect mechanophore signal. In contrast, several events localized within a single voxel produce a much stronger signal. Additionally, the simulations show that bond scission results in more microscopic motion in DNs as compared to SNs, see Fig.~\ref{fig:sim-dynamics-BB}c, which may also explain why in experiments the region over which activity is detected is much larger in DNs than in SNs. Note that enhanced dynamics entails more dissipation, thus providing a further explanation of the effectiveness of DNs in dissipating energy without catastrophically failing. 

Crucially, simulations allow the investigation of quantities not accessible in experiments, in particular stress redistribution across the matrix network to mitigate the localization of damage and to delay the onset of crack propagation. We found that in DNs a bond breaking event leads to a \textit{decrease} of the average load on the surrounding bonds and even an overall decrease of load when the scission occurs in the matrix (Fig.\ref{fig:sim-stress-dyn-upon-BB} c,d). This unexpected stress relaxation pattern is a key finding in explaining the exceptional toughness of DNs and underlines the value of 3D, molecular-level modelling.

Our work highlights that sacrificial bond breaking is necessary but not sufficient for toughness. Rather, it is the coupling of bond scission with efficient spatial stress redistribution associated with enhanced microscopic dynamics that fundamentally controls the fracture process in multiple-network elastomers. Our integrated experimental and numerical approach thus opens a pathway for rational design strategies targeting spatial control of stress dissipation, with broad implications for tough, damage-tolerant soft materials.

\section*{Materials and Methods}
\subsection*{Experiments}
\subsubsection*{Synthesis of the polymer networks}
Polymer network samples were synthesized using ethyl acrylate (EA), butanediol bis(acrylate) (BDA), 2-hydroxy-2-methylpropiophenone (HMP) as the monomer, cross-linker, and UV initiator, respectively. All chemicals were purchased from Sigma Aldrich; EA was purified over a column of activated alumina to remove the inhibitor before use, while the other chemicals were used as received. Melamine resin colloidal particles (MR) with $418~\mathrm{nm}$ diameter supplied from MicroParticles GmbH in powder form were used to obtain a suitable scattering signal (see details below). Synthesis was carried out in a glove box (Mbraun Unilab) under a nitrogen atmosphere and after bubbling all reactants  with nitrogen for a minimum of 30 minutes, to avoid side reactions with oxygen and water.

The synthesis follows the preparation procedure of Ref.~\cite{millereau2018mechanics}, but with the addition of $4 \times10^{-4} - 6 \times10^{-4}$ mass fraction MR particles to ensure an adequate single scattering signal with a transmission of $\approx 80 \%$ for all samples. The particle radius is 209 nm and the interparticle average distance is $\approx 5-20~~\qty{}{\micro\meter}$, both much larger than the mesh size, which is of the order of $\xi \approx 2$~nm, as estimated using $E \sim k_\mathrm{B} T/\xi^3$, with $k_\mathrm{B}$ Boltzmann's constant, $T$ absolute temperature and $E$ the Young modulus~\cite{rubinstein_polymer_2003}. 
For both SNs and DNs, addition of particles at these low volume fractions modestly increases the Young modulus, by about 16\% and 25\% for SNs and DNs, respectively, see Fig. S1 in the \SuppInf.
The particles were dispersed in EA through vortexing before stirring in an ultrasonic bath for 1h. Then, BDA and HMP were added, at molar concentrations of  $(0.5~\mathrm{mol} \%)$ and $(1.16~\mathrm{mol} \%)$ relative to the molar concentration of EA, respectively. All reactants were stirred for 5 minutes to obtain homogeneous precursor solutions, which were injected into a $1~\mathrm{mm}$ thick glass mold. Single networks were free radically polymerised after 2 hours of UV curing. 
Double networks were made by swelling single networks to equilibrium in a bath of monomer EA, $0.01~\mathrm{mol} \%$ of BDA and $0.01~\mathrm{mol} \%$ of HMP. The first (sacrificial) network thus swollen was removed from the bath, clamped between polyethylene terephthalate sheets, and pressed between glass plates. The polymerization of the second (matrix) network was induced by 2h of UV exposure. The final polymer networks had a refractive index of 1.47 and the pre-stretch of the sacrificial networks within them was $\lambda_0 = 1.7$ .   

The polymer sheets were cut to size using an Arbor press and die to yield $x = 50~\mathrm{mm}$ by $y = 10~\mathrm{mm}$ samples, see inset of Fig. \ref{fig:exp-stress-strain}b and Fig. S2, with thickness of about $0.9~\mathrm{mm}$ and $1.5~\mathrm{mm}$ for SNs and DNs, respectively. A $ 2~\mathrm{mm}$ notch parallel to the $y$ axis was made with a fresh razor blade, before mounting the sample on  the custom built rheometer described below. 

\subsubsection*{Experimental Setup}

The experimental setup is described in detail in \cite{orr_probing_2025a}, see Fig. SI2 in the \SuppInf~for a scheme. In brief, a laser beam with in-vacuo wavelength of 532 nm and typical power of 100 mW (Samba, by Cobolt) is spatially filtered, split into forward, FS, and backward, BS, scattering paths and shaped with lenses so as to illuminate a sample area of size up to $10 \times 20~\mathrm{mm}^2$ of the sample. Two shutters allow for alternating FS and BS illumination. Objective lenses form an image of the sample on the detectors of two CMOS cameras (Basler acA2000-340km Camera Link), using light scattered in the FS or BS direction, according to the shutters' status. With this photon correlation imaging (PCI) scheme~\cite{duri_resolving_2009}, each camera pixel corresponds to a well-defined location in the sample and receives light scattered at a well-defined scattering vector, with magnitude $q = {4 \pi n}{\lambda^{-1}} \sin\theta$, where $\theta$ is the scattering angle and $n = 1.47$ is the refractive index of the poly(ethyl acrylate) network. PCI probes the dynamics on a length scale $\sim 1/q$, projected on the direction of the scattering vector. The components of $\vec{q}$ are $[q_x,q_y,q_z] = [2.2, 1.9 \times 10^{-3},1.3 \times 10 ^{-1}]~\qty{}{\micro\meter^{-1}}$ for FS$_x$, $[q_x,q_y,q_z] = [4.9 \times 10^{-4}, 2.0, 7.0 \times 10 ^{-2}]~\qty{}{\micro\meter^{-1}}$ for FS$_y$, $[q_x,q_y,q_z] = [1.5, 8.3 \times 10^{-2}, 3.5 \times 10]~\qty{}{\micro\meter^{-1}}$ for BS (camera 1) and $[q_x,q_y,q_z] = [3.3 \times 10^{-1}, 1.5 \times 10^{-1}, 3.5 \times 10]~\qty{}{\micro\meter^{-1}}$ for BS (camera 2), where the sample lays in the $(x,y)$ plane and $x$ is the pulling direction, see Fig. SI2 in the \SuppInf. The activity signals from the two BS cameras were found to be the same. The average forms the reported BS signal.  

The poly-(ethyl acrylate) polymer networks are deformed simultaneously to the PCI experiments thanks to a custom-built, strain-controlled rheometer. The sample is clamped on both sides to holders that are displaced in opposite directions at the same speed, using precision stepper motors (LTA-HL by Newport). The extension speed $v$ is set to achieve an engineering strain rate $\Dot{\varepsilon} = \frac{v}{l_0} = 10^{-5}~\mathrm{s}^{-1}$ for samples of initial length $l_0 \approx 25~\mathrm{mm}$. The clamps displacement is monitored by a position detector (Keyence IL-065 laser head with an IL-1000 sensor). A force gauge (Andilog Centor Star Touch and SBlock force load cell) measures the resistance force $F$ exerted by the sample, from which the engineering stress is obtained as $\sigma = F/(l_0h_0)$, with $h_0=10~\mathrm{mm}$ the dimension of the unstretched sample along the $y$ direction, set by the size of the die. The precise value of $l_0$,  typically around $25~\mathrm{mm}$, is measured for each sample as the clamp distance where the force begins its up turn during extension. Scattering from imperfections on the sample surface is suppressed by immersing the clamps and sample in a custom-built bath filled with glycerol~\cite{orr_probing_2025a}, with refractive index $\approx 1.47$, very close to that of the networks. 

The crack tip trajectory is determined manually, by inspecting PCI movies with a time resolution of 2 s. The tip velocity is calculated from the change in position over two consecutive frames, using the calibration 1 pixel =  $\qty{0.025}{\micro\meter}$ determined by imaging a ruler at the sample position and subsequent analysis in ImageJ.

\subsubsection*{Photon Correlation Imaging Analysis}
The microscopic dynamics are quantified by $c_I$~\cite{duri_resolving_2009}, the degree of correlation between the intensity scattered at time $t$, corresponding to a stretch $\lambda(t) = l(t)/l_0$, and that scattered at a subsequent time $t+ \tau$, corresponding to a stretch $\lambda(t) + \Delta \lambda(\tau) = [l(t)+v\tau]/l_0$: 
\begin{equation}
    c_I(\lambda, \Delta \lambda, \vec{r}) = \beta \dfrac{\langle I_p(\lambda)I_p(\lambda + \Delta \lambda) \rangle_{\mathrm{ROI}(\vec{r})}}{\langle I_p(\lambda) \rangle _{\mathrm{ROI}(\vec{r})} \langle I_p(\lambda + \Delta \lambda) \rangle_{\mathrm{ROI}(\vec{r})}} - 1\,.
\end{equation}
Here, $\beta \lesssim 1$ is a normalization factor such that $c_I\rightarrow 1$ for $\Delta \lambda \rightarrow 0$, $I_p$ is the intensity of the $p$-th pixel, and the average $< \cdot\cdot\cdot >_{\mathrm{ROI}(\vec{r})}$ is performed over a region of interest (ROI) centered around position $\vec{r}$ in the sample, with size 0.8 by $0.8~\mathrm{mm}^2$ (32 by $32~\mathrm{px}^2$).

For a finite $\Delta \lambda$, $c_I < 1$ even in the linear elastic regime, due to the network affine deformation. In this regime, for a fixed $\Delta \lambda$, $c_I$ is constant with $\lambda$, with only small fluctuations. At larger $\lambda$, network damage results in fluctuations and an overall decrease of $c_I$. To focus on dynamics in excess to the linear affine deformation, we introduce a spatially- and temporally-resolved dynamic activity $A$, defined as
\begin{equation}
    A(\lambda, \Delta \lambda, \vec{r}) = \frac{\left | c_I(\lambda, \Delta \lambda, \vec{r}) - \langle c_I(\lambda, \Delta \lambda, \vec{r})\rangle_{\lambda^{lat}} \right|}{\langle c_I(\lambda, \Delta \lambda, \vec{r})\rangle_{\lambda^{lat}}}\,,
    \label{Activity}
\end{equation}
where $\lambda^{lat}$, is a latent stretching window before significant changes in $c_{I}$ occur. Typically, $\lambda^{lat}$ is centered around $\lambda \sim 1.08$ for SN and $\sim 1.4$ for DN and extends over a $900~\mathrm{s}$ time window, before any increase in $A$ is observed.   
We use $\Delta \lambda \approx 8\times 10^{-5}$ corresponding to $\tau = 8 ~\mathrm{s}$. This choice allows for good temporal resolution as we observe activity spikes in double networks lasting roughly $\approx 40~\mathrm{s}$ and crack propagation timescales of $\approx 70~\mathrm{s}$. Changing $\tau$ in the range $4~\mathrm{s}$ to $20~\mathrm{s}$ does not change the overall behavior reported in the main text. Values of $A \lesssim 0.8$, as typically observed here, correspond to microscopic displacements of order $\lesssim \qty{1}{\micro\meter}$ and $\lesssim 70~\mathrm{nm}$ for forward and backward scattering, respectively. 

\subsection*{Numerical simulations}

\subsubsection*{Numerical polymer model}
We use a coarse-grained approach in which polymeric strands are modeled as chains of bonded beads, following a methodology similar to the well-known Kremer-Grest model~\cite{grest1986molecular, everaers2020kremer}. Beads interact with each other through a repulsive potential (to avoid overlaps) and an attractive potential to form polymeric chains and to create cross-links. Bending rigidity is accounted for by adding a three-body angular potential between bonded beads.
The repulsive  potential between beads separated by a distance $r$ is modelled by the Weeks-Chandler-Andersen (WCA) potential, which consists of a truncated and shifted Lennard-Jones (LJ) potential :
\begin{equation}\label{eq_imp_LJ}
    U_\mathrm{WCA}(r) = \begin{cases}
    4\varepsilon \left[ \left(\dfrac{\sigma}{r}\right)^{12} - \left(\dfrac{\sigma}{r}\right)^{6}\right] + \delta, & r\leq r_c \\
     0 , & r > r_c
    \end{cases}
\end{equation}
where $\varepsilon$ is the interaction strength, $\sigma$ represents the diameter of a particle, $r_c  = 2^{1/6} \sigma$ is the potential cutoff distance, and $\delta = \varepsilon$ ensures continuity at $r = r_c$. All results are reported in reduced units using the bead size $\sigma$, the typical interaction energy $\varepsilon$, and the particle mass $m$.
The reduced time unit we use is $\tau = \sqrt{\frac{\varepsilon}{m\sigma^2}}$ (the typical vibrational time of beads) and the stress unit is $\frac{\varepsilon}{\sigma^3}$ (the typical force between beads per bead size area). It is possible to re-scale time and stress knowing the typical bead size (about 3 monomers) and mass together with the typical energy scale of bead interactions to obtain comparable units with respect to the experiments \cite{everaers2020kremer}.


During polymer network synthesis, adjacent beads within a chain are connected via an attractive finite extensible nonlinear elastic (FENE) potential, given by:
\begin{equation}
    U_\mathrm{FENE}(r) = \begin{cases}
    -\frac{1}{2} k R_0^2 \ln\left[ 1-\left(\dfrac{r}{R_0}\right)^2\right], & r \leq R_0\\
     \infty , & r > R_0
    \end{cases}
    \label{eq-fene-potential}
\end{equation}
where $k=30 (\varepsilon/\sigma^2)$ and $R_0 = 1.5 \ \sigma$. This specific choice prevents chains from crossing each other~\cite{grest1986molecular,murat1999statics,kremer1990dynamics}. 

At the end of the synthesis protocol, the FENE bond potential is replaced by the following quartic potential to allow for bond scission, when the bond length exceeds a length $R_0$ during uni-axial extension tests:
\begin{eqnarray}
U_\mathrm{bond}(r) = K (r-R_c)^3(r-R_c - B) + U_0 + U_{WCA}(r)  
\label{bond_break}
\end{eqnarray}
where $K = 2351$, $R_c = R_0$, $U_0 = 92.74467$, $B = -0.7425$. 
This choice of parameters prevents the chains still from crossing each other and has the same equilibrium length as the FENE potential \cite{ge2013molecular}. 

In our simulations, bonded beads also interact via a three-body angular potential to account for bending rigidity:
\begin{eqnarray}
U_\mathrm{angular}(\theta) = K_\theta \left ( 1 + \cos{\theta} \right )  
\label{eq-angular-potential}
\end{eqnarray}
where $\theta$ is the angle between three consecutive monomers along the polymer chain and $K_\theta=1.276$ is chosen to mimic the bending rigidity of poly(ethyl acrylate) (PEA) polymers~\cite{millereau2018mechanics,everaers2020kremer}. We adjust the density at which the polymerization and cross-linking are performed to the experimental synthesis protocol. It is known that the response is sensitive to network preparation and architecture~\cite{tauber2021sharing, tian2025influence}. 

\subsubsection*{Protocol for forming simulated double network samples}
%
\paragraph*{Sacrificial network}
A mixture of $N^\mathrm{beads}_1$ monomers of the sacrificial network, dimer cross-linker molecules (made of two bonded beads) with a concentration $c_1$,
and $N^\mathrm{radical}_1$ radical beads (with $N^\mathrm{radical}_1=N^\mathrm{beads}_1 c^\mathrm{r}_1$) is prepared at an overall density $\rho^\mathrm{0}=0.8$.
We use a cross-link concentration $c_1=5\%$ and a radical concentrations $c^\mathrm{r}_1 = 1/(200 \sigma)$ (such that the chain length resulting from the radical-like polymerization in the absence of cross-linking would be $\ell_1 = 200 \sigma$).
All beads interact via the same repulsive WCA potential.
Radical-like polymerization and cross-linking are then performed, creating FENE bonds between bonded atoms, leading to a single network of size $L\times L \times L$ (Fig.~\ref{fig:exp-stress-strain}-a), left panel). 
The network thus formed is heterogeneous and the strand length between two consecutive cross-links follows an exponential distribution \cite{flory1953principles, rubinstein2003polymer}.

\paragraph*{Swelling and matrix network}

$N^\mathrm{monomers}_2$ matrix beads, cross-linker molecules at a concentration $c_2=1\%$ for the matrix network and $N^\mathrm{radical}_2$ radical beads
($N^\mathrm{radical}_2=N^\mathrm{beads}_2 c^\mathrm{r}_2$ with $c^\mathrm{r}_2 = 1/(1000 \sigma)$ such that $\ell_2 = 1000 \sigma$) are added.
The repulsive potential between the beads of the second network is first set as a ``soft potential": $E=A\left(1+\cos\pi\frac{r}{r_c} \right )$ for which the prefactor $A$ is slowly increased in time (ramped from 0 to 50 during 100 $\tau$) in order to separate progressively overlapping atoms in the NPT ensemble (with a pressure $P=3.8 \epsilon\sigma^{-3}$ in reduced LJ units). The potential is switched to WCA and the sample is relaxed in the NVT ensemble with a sample size $\lambda_0 L\times \lambda_0 L \times \lambda_0 L$ (Fig.~\ref{fig:exp-stress-strain}-a), middle panel).
Simultaneous radical-like polymerization and cross-linking is then performed (by creating FENE bonds between bonded atoms (Eq.~\ref{eq-fene-potential})) to obtain a heterogeneous double network of size $\lambda_0 L\times \lambda_0 L \times \lambda_0 L$ (Fig.~\ref{fig:exp-stress-strain}-a, right panel).
Finally, the FENE bonds are switched to quartic bonds and the system is relaxed ($R_0=1.5$).

The total number of beads of the network is $N = 440 000$.
The system size is $\lambda_0 L=82 \sigma$, with $\lambda_0 = 1$ for the SN and $\lambda_0 = 2$ for the DN. The corresponding mesh sizes for the sacrificial and the matrix network are $10 \sigma$ and $50 \sigma$.

Results averaged over two single network and three double network samples are shown.

\subsubsection*{Stress calculation}
The stress response is calculated from both the virial stress (which has contributions from the pair, bond and angular potential) and the kinetic contribution (although the kinetic contribution is small) \cite{thompson2009general}. The true stress is calculated as $\sigma_{T} = \sigma_{xx} - \frac{1}{2}(\sigma_{yy} + \sigma_{zz})$.
The engineering stress reads $\sigma_e = \sigma_T/\lambda $.

Stress maps are obtained by outputing a per-atom stress tensor for each atom, and then averaging over a small volume $dx dy dz$ with $dx=5\sigma$ and $dy=dz=2\sigma$.

\subsubsection*{Pressure-controlled uni-axial deformation protocol}
Uni-axial stretching of the simulation box is performed using a step-wise protocol \cite{bouzid2018network} in which the box dimensions and particle positions are affinely displaced by $\Delta \lambda  = 1\%$ along the $x$-axis at each step. After each deformation step, the system is relaxed for a period of time $t_{relax}$ under a Langevin dynamics at temperature $T=1.0$, with a damping coefficient $\xi = 1.0$, ensuring a strain rate of $\dot \lambda=\Delta \lambda/t_{relax} = 4 \cdot 10 ^{-5} \tau ^{-1}$ . A Berendsen barostat is applied in the $y$ and $z$ directions (uncoupled) to maintain a constant pressure of $P = 3.8$, corresponding to the equilibrium pressure prior to deformation. Periodic boundary conditions are applied in the simulation box. The integration time step is set to $\Delta t = 0.005$. %
Simulations are performed using the LAMMPS
package \cite{thompson2022lammps} and visualization is performed through the OVITO software \cite{stukowski2009visualization}.

\subsubsection*{Average response to single bond breaking events}
In order to compare typical network configurations before and after the occurrence of a bond breaking event, we average each configuration over (at least) 150 snapshots spaced by $0.2 \tau$ -  $2 \tau$, leading to an averaging time window $T_\mathrm{av}$ ranging between  $30 \tau$ -  $300 \tau$
(see the \SuppInf~and Fig. S10 for the detailed protocol). The average displacement field and change in stress due to a bond breaking event is calculated using these averaged configurations.

The magnitude of the average displacement between two consecutive configurations averaged over a time windows $T_\mathrm{av}$ 
is however non-zero even in the absence of bond breaking, because of the slow relaxation dynamics of long polymer strands (mean square displacement reaching a plateau after $\sim 1000 \tau$), which is inherent to the non-equilibrated nature of our networks in a large-deformation protocol at finite strain rate. We checked that the baseline displacement measured far from a bond breaking event is consistent with the displacement measured in the absence of a rupture event (see Fig. S12 in the \SuppInf), such that this background contribution could be safely removed.

Bond breaking events induce a stress response in the surrounding strands that can be seen in the pair, bond and angular contributions to the interaction potential (see Fig. 11 in the \SuppInf). The dominant contribution is due to the bond interaction potential.

The maps of displacement and stress change shown in Fig.~\ref{fig:sim-stress-dyn-upon-BB}a,c are obtained by integrating along $y$ over the sample thickness and averaging in the $(x,z)$ plane using bins of size $dx = 5\sigma$, $dz=2\sigma$.
The graphs of Fig.~\ref{fig:sim-stress-dyn-upon-BB}b,d are obtained by azimuthally averaging data in a slice of thickness $dx= \sigma$ centered around $x=0$ and parallel to the $(y,z)$ plane.

\subsubsection*{Density fluctuations}
Density profiles, $\rho(x)$, along the stretching direction $x$ are computed by integrating the local density along $y$ and $z$, the directions perpendicular to the stretching direction, and then by binning along $x$ using a bin width $\Delta x = 10$ (see Fig. S7 in the \SuppInf).

The density localization is computed as $\Delta \tilde{\rho} = (\rho_\mathrm{max} - \rho_\mathrm{min})/\rho_{0}$ with $\rho_\mathrm{min}$ and $\rho_\mathrm{max}$ the minimum and maximum densities in the profile, respectively, and $\rho_{0}$ the average density. A density localization larger than 1 means that the network has one or several regions with a density $\rho_\mathrm{min} \simeq 0$ (fractured regions), with other regions being densified after fracture, yielding $\rho_\mathrm{max}>\rho_0$.

\subsubsection*{Overlap parameter}
We compute the overlap from the cross-link positions between two instantaneous configurations $\alpha$ and $\gamma$ (i.e., not averaged over thermal fluctuations) as:
    \begin{equation}
        Q (\mathbf{r_\alpha},\mathbf{r_\gamma}) = \frac{1}{N_c}\sum_{i} \omega\left (\frac{| \mathbf{r}_{\alpha,i}-\mathbf{r}_{\gamma,i} |}{a}\right) \,
    \end{equation}
where the sum extends over the $N_c$ simulated cross-link beads, $a$ is a cutoff parameter and $\omega(x)$ a step function, $\omega = 1$ for $x<1$ and $\omega = 0$ for $x \geqslant 1$. The cutoff parameter is chosen by inspecting the probability distribution of the bead displacements, which contains contributions from both thermal motion and strain-induced rearrangements, see Fig. S9 in the \SuppInf. We choose $a=10$, large enough to exclude most of the small fluctuating displacements due to thermal motion, but still small enough to be sensitive to differences in displacement as a function of stretch $\lambda$ and to preserve good statistics. Moreover, focusing only on cross-link bead positions enables one to accelerate the overlap computation and remove some noise due to the large motion of beads within the strand, espcially at low stretch.

\section*{Supplementary Material}
\subsection*{Experiments}
\subsubsection*{Effect of inclusion of melamine resin colloidal particles on the Young modulus of the polymer networks}

We add melamine resin colloidal particles at a $4 \times10^{-4} - 6 \times10^{-4}$ mass fraction, in order to optimize the scattering signal. As shown in Fig.~\ref{Fig:PureParticles}, this results in a modest increase of the Young's modulus, of about 16\% for SNs and 25\% for DNs, to be compared to a typical sample-to-sample variability of about 15\% and 16\% for SNs and DNs, respectively. The Young moduli for single and double networks with particles are $0.84 \pm 0.13$~MPa and $1.27 \pm 0.20$~MPa, respectively. 

\setcounter{figure}{0}
\makeatletter
\renewcommand{\thefigure}{S\@arabic\c@figure}
\makeatother
\begin{figure}[H]
\centering
\includegraphics[width=1\columnwidth]{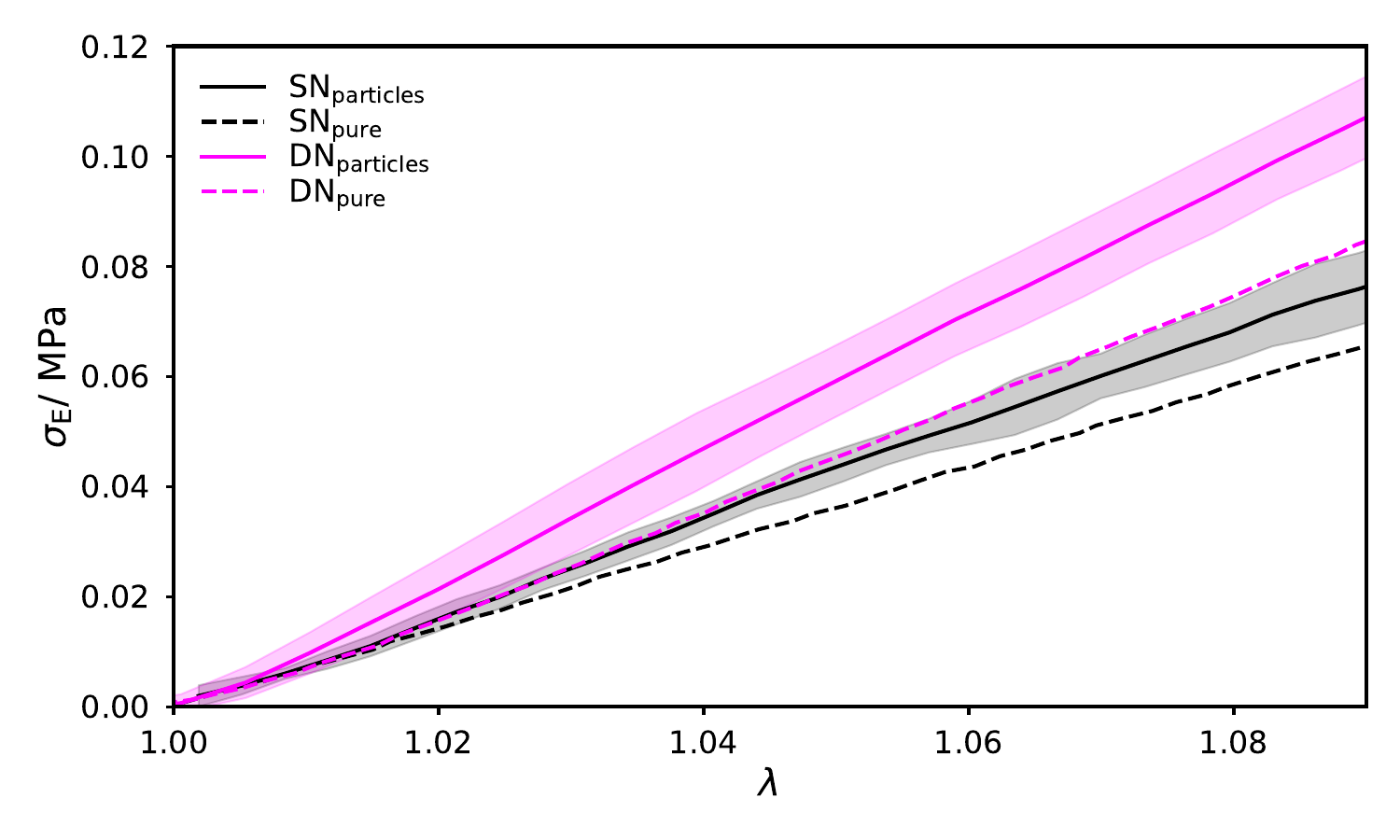}
\caption{Engineering stress $\sigma_\mathrm{E}$ in response to stretching $\lambda = L/L_0$ for single and double polymer networks with and without colloidal particles. Data for samples with particles are averaged over five SNs and three DNs, respectively. The shaded areas show the standard deviation over measurements on distinct networks of the same kind.}
\label{Fig:PureParticles}
\end{figure}

\subsubsection*{Experimental setup}
\begin{figure}[H]
\centering
\includegraphics[width=0.5\textwidth]{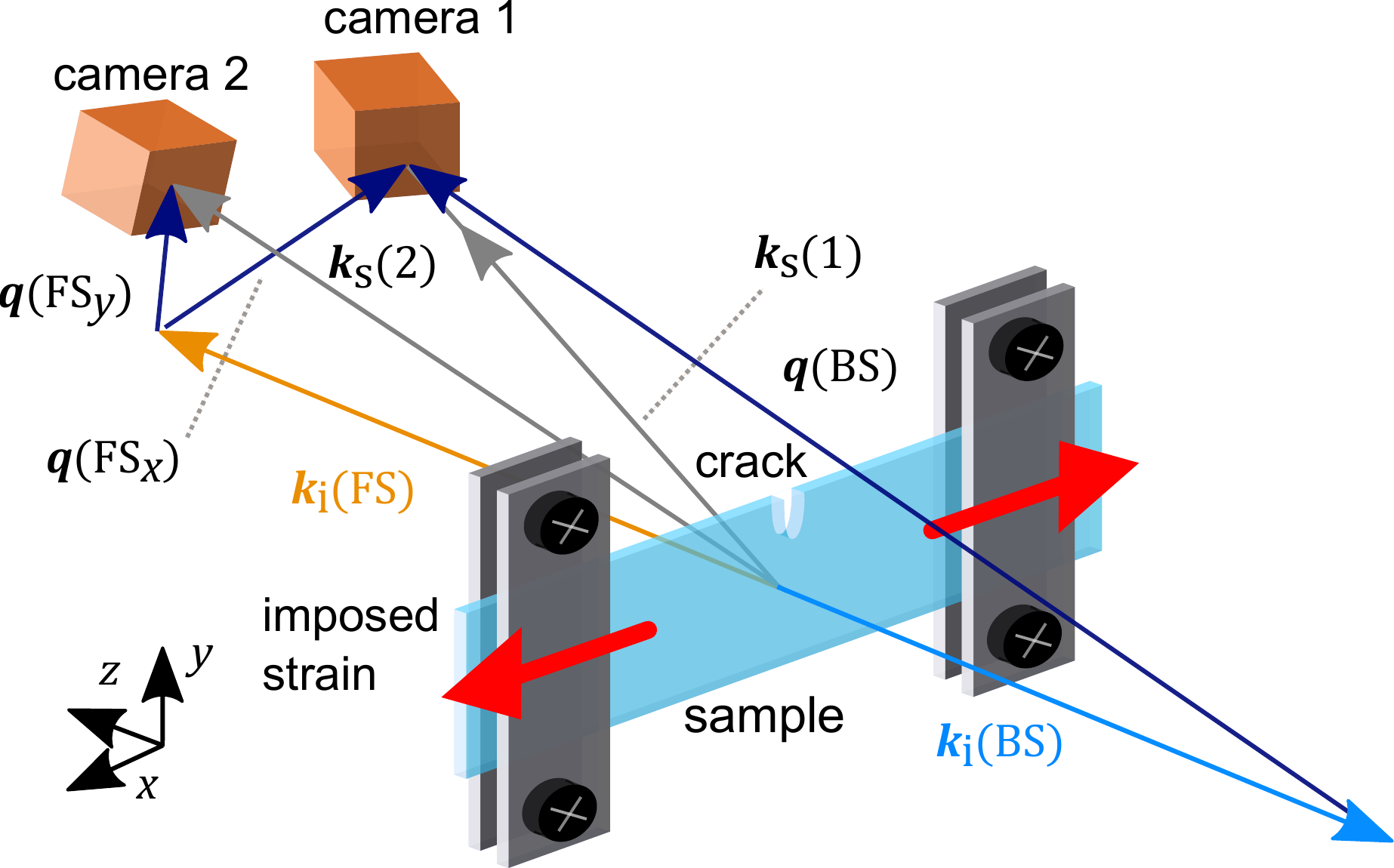}
\caption{Scheme of the experimental setup. The sample is held by two clamps fixed to motors (not shown) to apply uni-axial elongation. The axis of one motor is equipped with a force sensor (not shown), to measure the stress. Laser light illuminates the sample alternating between the directions of the wave vectors $\vec{k}_i(FS)$ and $\vec{k}_i(BS)$, for forward and backward scattering, respectively. Two CMOS cameras take images of the sample formed by scattered light, either in the scattering plane $(x,z)$ (camera 1), or slightly above it (camera 2). $\vec{k}_\mathrm{S}(1)$ and $\vec{k}_\mathrm{S}(2)$ are the wave vectors of scattered light collected by camera 1 and 2, respectively. The scattering vectors for the BS and FSx, FSy geometries are indicated by the blue arrows and $\vec{q}$ letter.
}
\label{Fig:setup}
\end{figure}

Figure \ref{Fig:setup} shows schematically the setup coupling tensile tests and Photon Correlation Imaging~\cite{Duri2009}. See Ref.~\cite{orr_probing_2025a}
and the \MM~section of the main text for details. 

\begin{table}[H]
\centering
\caption{Samples used in distinct fracture experiments for which data are shown in the figures of the main text. All samples of the same kind (SN or DN) have the same composition. Figures \ref{fig:SN_repeat}-\ref{fig:spatial_activity} show the results for all samples.}
\label{table_repeats}
\begin{tabular}{lrrrrrrr}
Main text figure & SN I & SN II & SN III & DN I & DN II & DN III & DN IV \\
\midrule
1b. &   \checkmark &  &   &   &   &   &  \checkmark \\
2. &  \checkmark &   &   &  \checkmark &   &   &   \\
3. &  \checkmark &   &   &   &   &   &  \checkmark \\
4. &  \checkmark &  \checkmark &  \checkmark &  \checkmark &  \checkmark &  \checkmark &  \checkmark \\
5. &  \checkmark &  \checkmark &  \checkmark &  \checkmark &  \checkmark &  \checkmark &  \checkmark \\
\bottomrule
\end{tabular}
\end{table}

\subsubsection*{Activity and notch speed during fracture}
\begin{figure}[H]
\centering
\includegraphics[width=1\columnwidth]{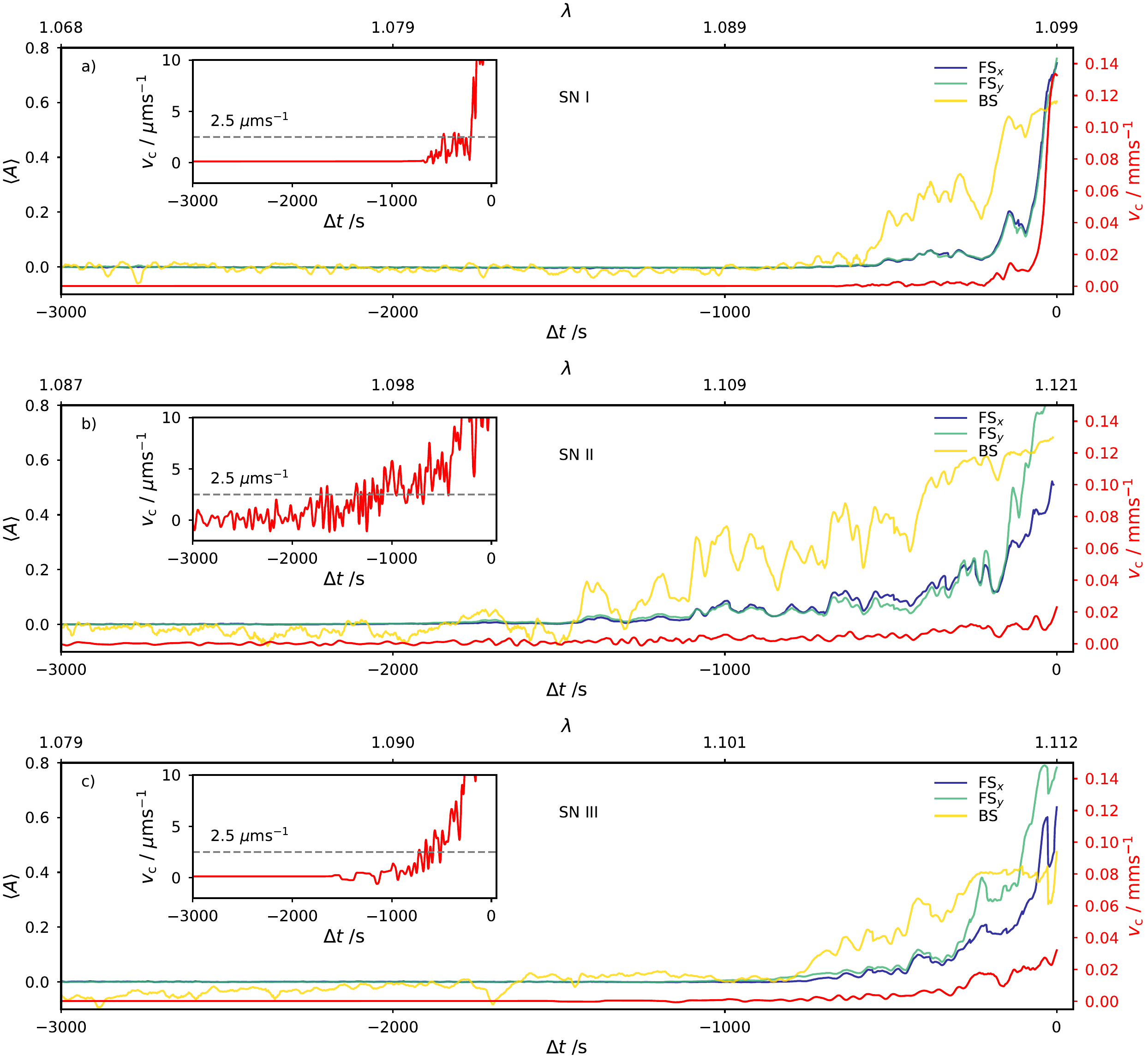}\caption{\textbf{Crack speed and microscopic dynamics prior to rupture for repeats of SN experiments.} The left axis shows time dependence of the microscopic rearrangements as quantified by the activity, $<A>$ (see main text for definition), averaged over the half plane ahead of the crack tip, for scattering in the forward (FS$_x$, FS$_y$) and backward (BS) directions. $\Delta t$ corresponds to the time to macroscopic rupture. The right axis and red line show $v_\mathrm{c}$, the propagation speed of the crack tip, with zooms on the behavior right before rupture in the insets. The dotted line in the insets is the threshold below which $v_\mathrm{c}$ is considered to be negligible for the analysis of Fig. \ref{fig:spatial_activity}.}
\label{fig:SN_repeat}
\end{figure}

\begin{figure}[H]
\centering
\includegraphics[width=1\columnwidth]{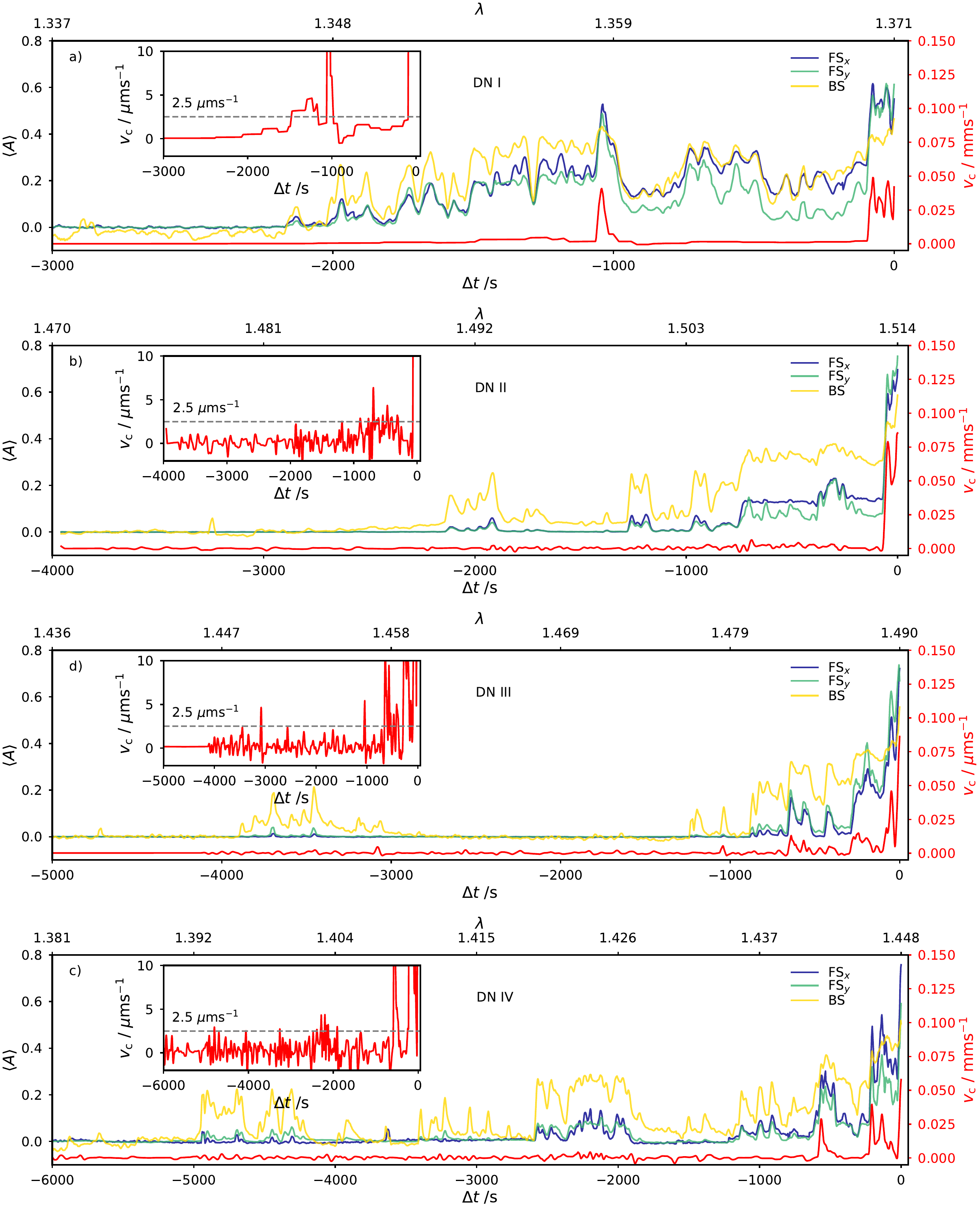}
\caption{\textbf{Crack speed and microscopic dynamics prior to rupture for repeats of DN experiments.} The left axis shows time dependence of the microscopic rearrangements as quantified by the activity, $\langle A \rangle>$ (see main text for definition), averaged over the half plane ahead of the crack tip, for scattering in the forward (FS$_x$, FS$_y$) and backward (BS) directions. $\Delta t$ corresponds to the time to macroscopic rupture. The right axis and red line show $v_\mathrm{c}$, the propagation speed of the crack tip, with zooms on the behavior right before rupture in the insets. The dotted line in the insets is the threshold below which $v_\mathrm{c}$ is considered to be negligible for the analysis of Fig. \ref{fig:spatial_activity}.}
\label{fig:DN_repeat}
\end{figure}

In the main text we show two representative experiments that highlight how activity in the sample volume ahead of the notch tip behaves as the samples are stretched (Fig. 3 of the main text). These data where collected for samples SN I and DN IV, see Table~\ref{table_repeats} that recaps all the studied samples. Figures~\ref{fig:SN_repeat} and~\ref{fig:DN_repeat} include repeat experiments for all SNs and DNs, respectively. From Fig.~\ref{fig:SN_repeat} we see that overall the SN experiments behave similarly, with increased activity directly before fracture. Additionally, SNs fracture at similar stretches, $\lambda \sim 1.10-1.12$.  Figure \ref{fig:DN_repeat} shows that intermittent bursts of activity and crack advancement are consistently exhibited by each DN, albeit with some differences, e.g. in the duration of the quiescent periods between bursts of activity. Interestingly, in DNs the sample-to-sample variations of the stretch at break are larger than for SNs ($1.4 \lesssim \lambda \lesssim 1.5$), consistent with the larger heterogeneity of the former, as discussed in the main text.

\begin{figure}[H]
\centering
\includegraphics[width=1\columnwidth]{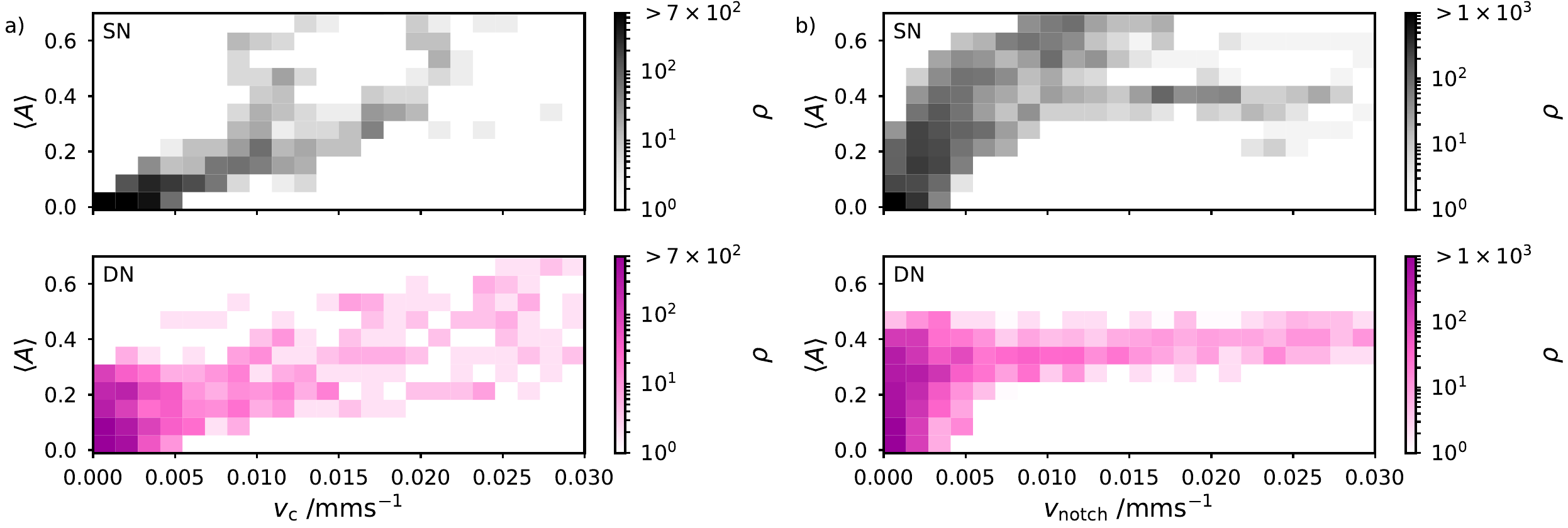}
\caption{\textbf{Correlation between crack propagation and microscopic dynamics depends on network architecture.} Color-coded number of occurrences of ($v_\mathrm{c}$, $\left<A\right>$) pairs obtained by binning measurements as those shown in Fig. 3 of the main text, for SNs (top) and DNs (bottom). The results from scattering vectors FS$_y$ and BS are shown in a) ad b), respectively. Data obtained from tests on three SN and four DN samples in the 3000~s preceding macroscopic failure. The excess of large values of activity at vanishing $v_\mathrm{c}$ in DNs as compared to SNs is clearly visible for the FS$_y$ data probing dynamics on a length scale $\lesssim1~\mu$m, while this feature is less evident for the BS data, which probe dynamics on shorter length scales ($\lesssim 70~\mathrm{nm}$), as discussed in the main text. }
\label{fig:2D_histogram}
\end{figure}

Figure~\ref{fig:2D_histogram} shows the 2D histogram of the joint occurrence of pairs $(v_c,\langle A \rangle)$ (crack speed and average activity in the half-plane ahead of the crack tip, respectively) for data collected in the FS$_y$ and BS scattering geometry. For the FS$_y$ geometry (left column of Fig.~\ref{fig:2D_histogram}) the overall behavior is similar to that of the FS$_x$ geometry discussed in  Fig. 4 of the main text: The crack speed and activity are correlated for SNs, displaying no activity when there is no crack propagation, while for DNs there is significant activity at low crack propagation speeds. By contrast, for the BS geometry (right column of Fig.~\ref{fig:2D_histogram}), which is sensitive to microscopic motion on smaller length scales $\sim 70$~nm, we observe activity when the notch is not propagating for both SNs and for DNs. As discussed in the main text, this points to the relaxation of density fluctuations on micrometric length scales as a key mechanisms for energy dissipation in DNs, not present in SNs.

\begin{figure}[H]
\centering
\includegraphics[width=1\columnwidth]{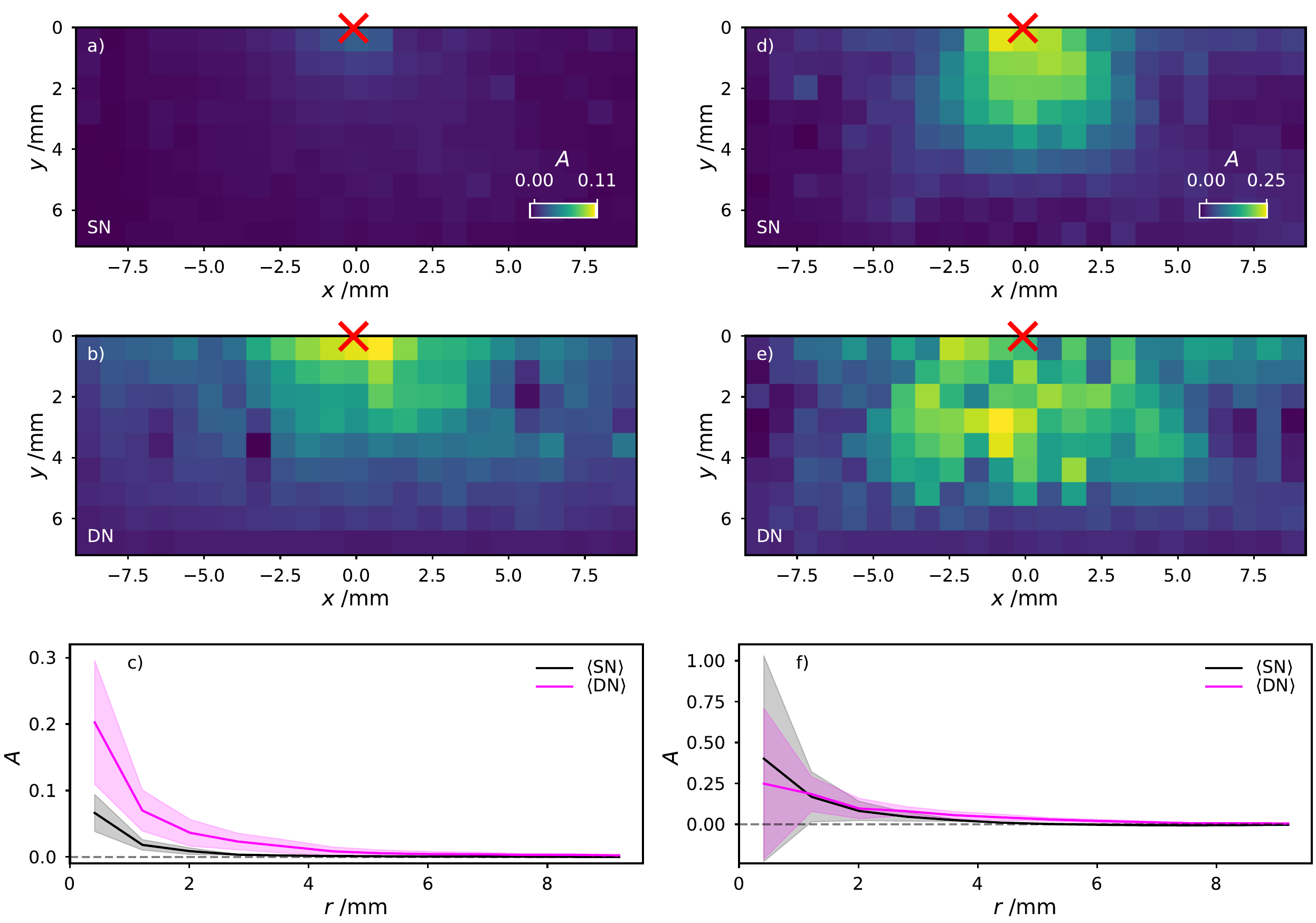}
\caption{\textbf{Spatial dependence of the activity in the half plane ahead the crack tip}. The results for the scattering vector FS$_y$ are shown in (a) through (c). Spatial maps of $A$, for SNs (a) and DNs (b), in a moving reference frame where the crack tip corresponds to $(x,y) = (0,0)$ (red cross). Data are obtained by averaging FS$_y$ measurements for all samples, in time intervals with negligible notch propagation speed, $v_c \leq \qty{2.5}{\micro\meter\second^{-1}}$, and within 3000 s before macroscopic rupture. (c) Activity as a function of distance $r$ to the crack tip, obtained by azimuthally averaging the data of a) and b). The shaded areas show the standard deviation resulting from sample to sample variations. The results for the scattering vector BS are shown in (d) through (f).}
\label{fig:spatial_activity}
\end{figure}

Similarly, we find that the spatial dependence of the activity in the half place ahead of the crack tip as measured from the FS$_y$ data follows the same trend as for the FS$_x$ data, compare Fig.~\ref{fig:spatial_activity}a-c to Fig. 5 of the main text. This has to be contrasted to the behavior of the short length-scales dynamics (BS data, Fig.~\ref{fig:spatial_activity}d-f), for which once again the differences between SNs and DNs are significantly less marked than for the micrometer-scale dynamics.

\newpage


\subsection*{Simulations} 

\subsubsection{Profiles of density fluctuations and bond breaking in regime II(a)}
Density profiles (black lines and small circles in Fig. S7) are obtained by integrating the number of atoms over the system thickness in the $y$ and $z$ directions and over slices of width $\Delta x = 10 \sigma$ in the stretching direction.

\begin{figure}[H]
\centering
\includegraphics[width=1\columnwidth]{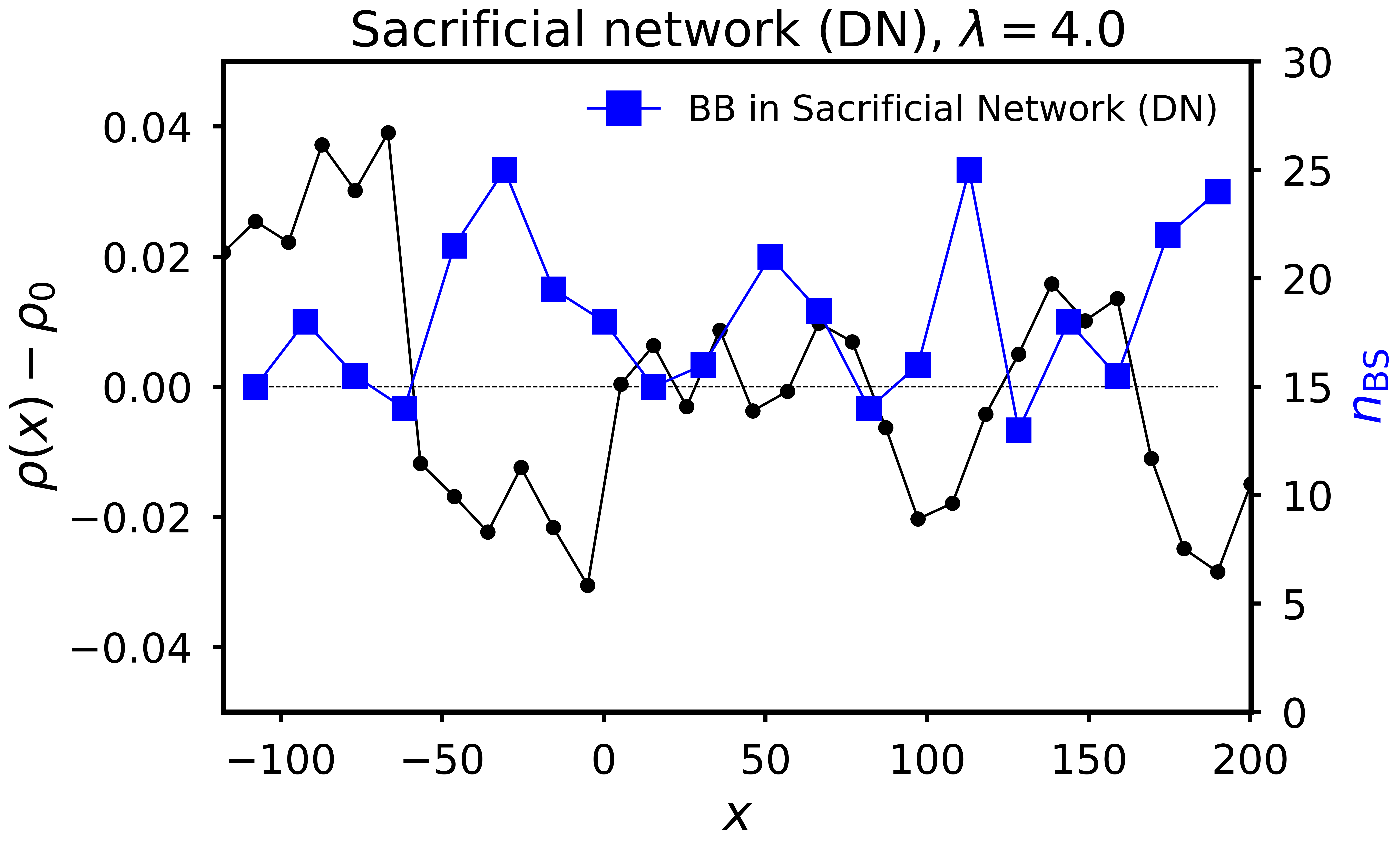}
\caption{Example of profile of density fluctuations in the (full) double network (black dots and solid line) with the corresponding broken bonds in the sacrificial (blue) and matrix (red) networks at the end of regime (ii).}
\label{fig:SI-profile-1}
\end{figure}

Similarly, profiles of bond breaking are obtained by integrating the number of bond breaking events over the system thickness in the $y$ and $z$ directions and over slices of width $\Delta x = 10 \sigma$ in the stretching direction.
At a stretch $\lambda = 4$, the matrix is still intact and all the events occur in the sacrificial network.

\subsubsection{Profiles of bond breaking in regime II(b)}
\begin{figure}[H]
\centering
\includegraphics[width=1\columnwidth]{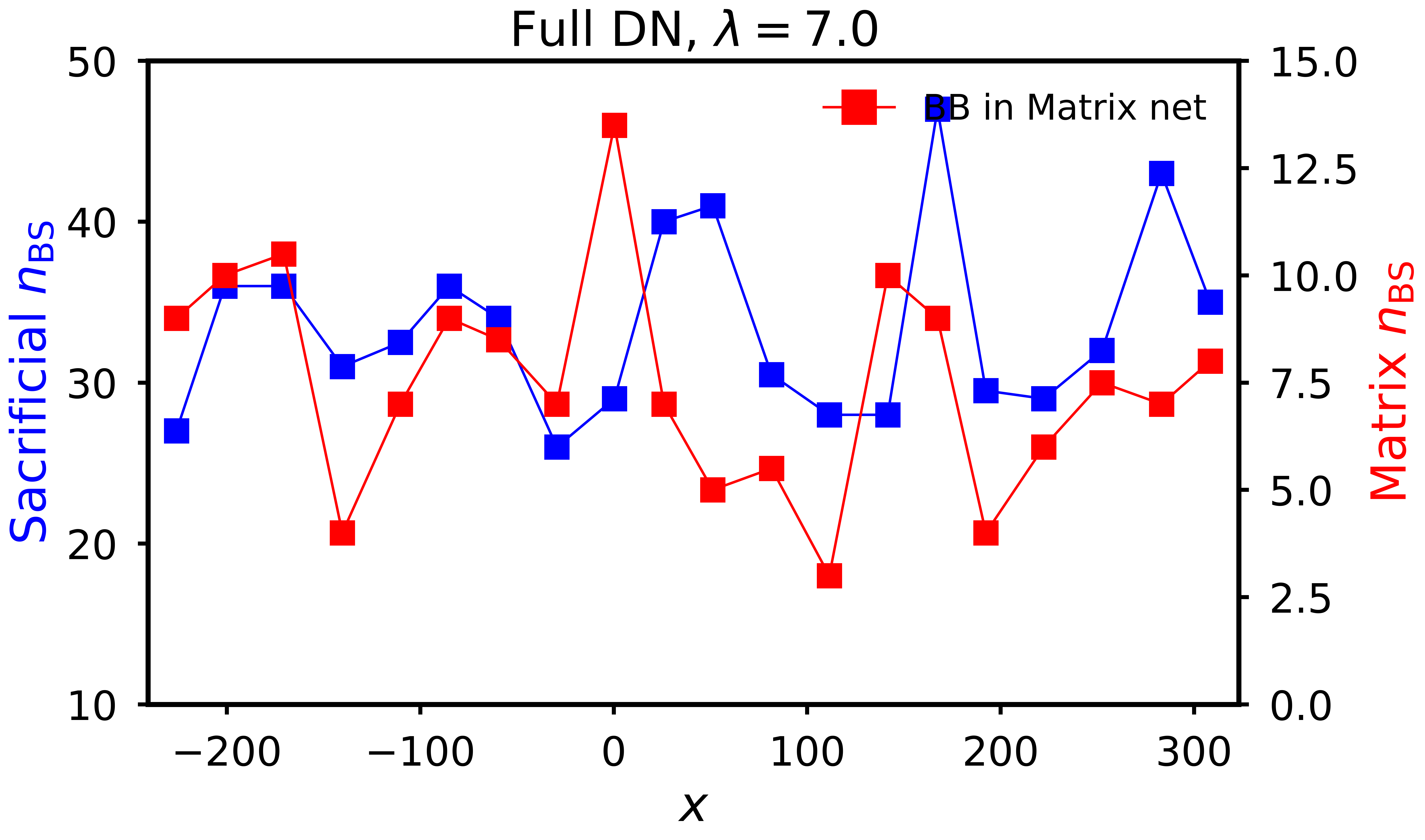}
\caption{Example of profile of density fluctuations in the sacrificial network embedded in the DN with the corresponding broken bonds in the sacrificial networks (blue) in regime (ii).}
\label{fig:SI-profile-2}
\end{figure}

At larger stretch values ($\lambda = 7$), rupture events occur in both networks.
Figure~\ref{fig:SI-profile-2} shows profiles of bond breaking in the matrix and sacrificial network, obtained by integrating the number of bond breaking events occurring in each network respectively over the system thickness in the $y$ and $z$ directions and over slices of width $\Delta x = 10 \sigma$ in the stretching direction.

\subsubsection{Overlap parameter}
Figure S9 depicts the distribution of displacement magnitude between two instantaneous network snapshots (no averaging).
Most atoms display a small displacement due to thermal noise. In order to filter some of this contribution, we compute an overlap parameter with a cutoff $a=10$ which enables us to focus on the large tails of the distributions, where bond breaking events contribute the most.
\begin{figure}[H]
\centering
\includegraphics[width=1\columnwidth]{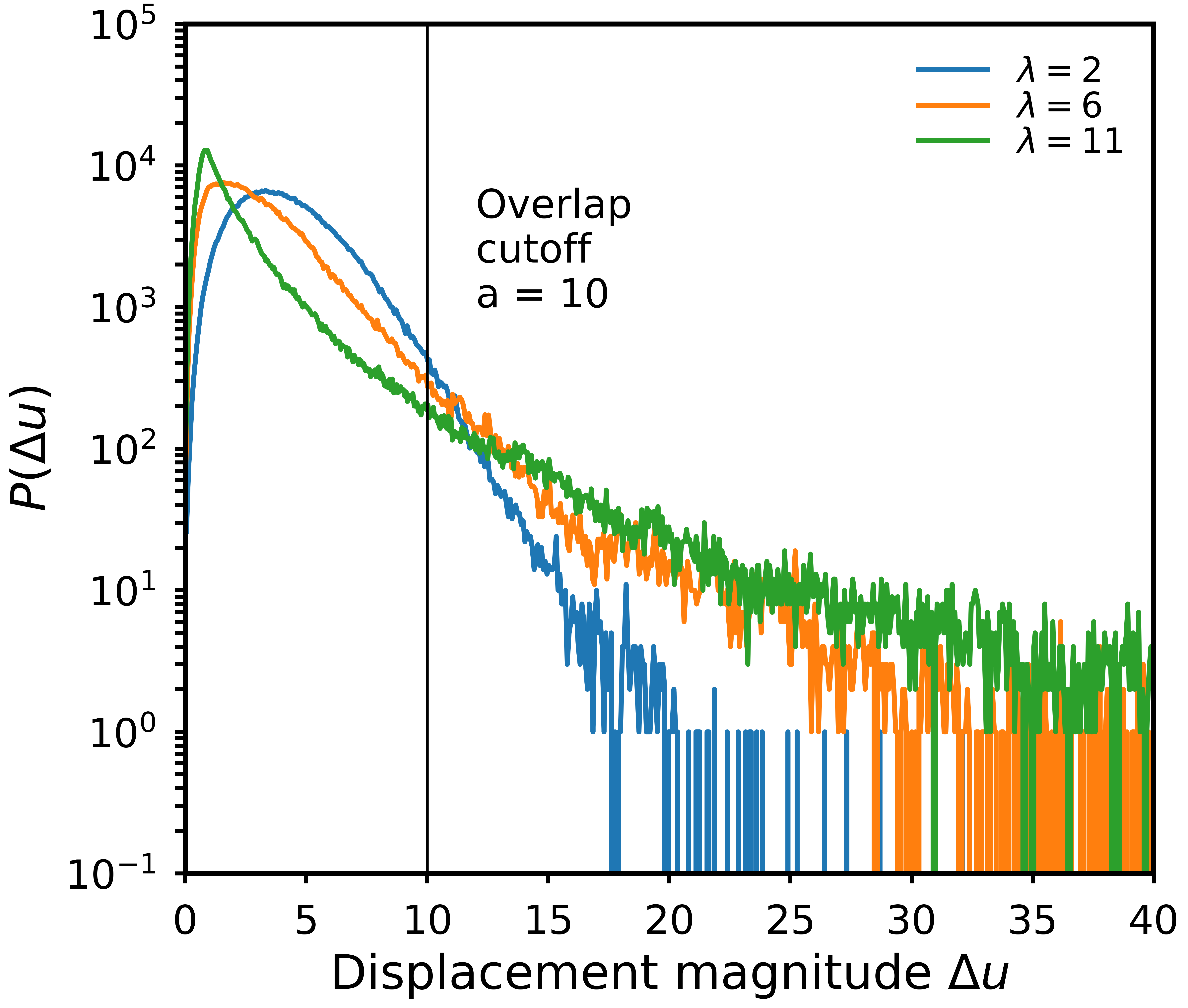}
\caption{Distribution of displacement magnitude between successive DN configurations at different stretch values $\lambda$. The black vertical solid line indicate the cutoff value chosen to compute the overlap.}
\label{fig:SI-disp_distrib}
\end{figure}

\subsubsection{Protocol to measure the average response to a single bond breaking event}

Fig. 9 of the main text depicts the average displacement and stress responses in response to a single bond breaking event. At the macroscopic scale, a bond breaking event leads to energy dissipation (as shown by the drop in potential energy in Fig. S10, left panel).

\begin{figure}[H]
\centering
\includegraphics[width=1\columnwidth]{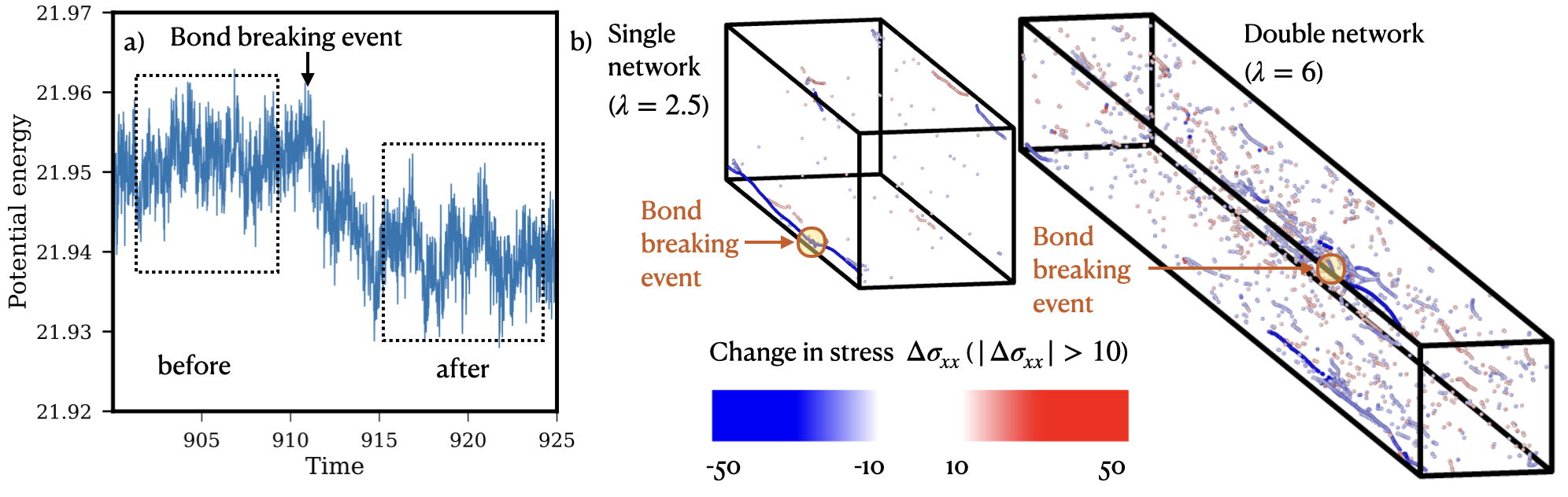}
\caption{Protocol to measure average response to single bond breaking events. a) Total potential energy of a single network sample as a function of time (in LJ units). b) Snapshots of single (left) and double (right) network configurations. Particles are colored depending upon the change of stress they experience between their average state before and after a bond breaking event. Only stress changes larger than 10 (in LJ units) are shown for clarity.}
\end{figure}

Since thermal motion induces significant particle displacement and stress and energy fluctuations even in the absence of bond breaking, we average the atoms' positions and the stress over several instantaneous snapshots sampled before and after the bond breaking event (see Fig. S10a) in order to obtain the response (displacement and change in stress) by comparing typical configurations.
Network configurations are saved every $0.2\tau$ or $2\tau$ (depending upon the rate of bond breaking: typically we take $2\tau$ for the DN and $0.2\tau$ for the SN), before and after the bond breaking event.
Typical configurations before and after a bond breaking event are obtained by averaging over at least 150 snapshots.
The $xx$ component of the stress change is computed as : $\Delta \sigma_{xx} = \sigma_{xx}^\mathrm{after} - \sigma_{xx}^\mathrm{before}$.
The displacement is computed as : $\Delta u_i = x_i^\mathrm{after} - x_i^\mathrm{before}$ for each component $i$.
The stress change in response to a single BB event is depicted for a SN (see Fig. S10b, left panel) and a DN (see Fig. S10b, right panel).
In order to obtain the average maps shown in Fig. 9 of the main text, we then average the response over several bond breaking events, namely : 25 events for the SN, 89 events for bond breaking events occurring the sacrificial network of the DN and 50 events occurring in the matrix network in the DN.
All these events are spread across the different regimes of deformation (see main text).

\subsubsection{Analysis of the different stress contributions}

Figure S11a depicts the different contributions of the interaction potential to the stress-strain curve, and one can see that the macroscopic stress response is dominated by the bond interactions. This is also true for the change in stress in response to bond breaking, which is dominated by a change in bond stress $\Delta \sigma_{xx}$ that can be positive, implying enhanced loading, or negative, corresponding to stress relaxation (Fig. S11b).

\begin{figure}[H]
\centering
\includegraphics[width=1\columnwidth]{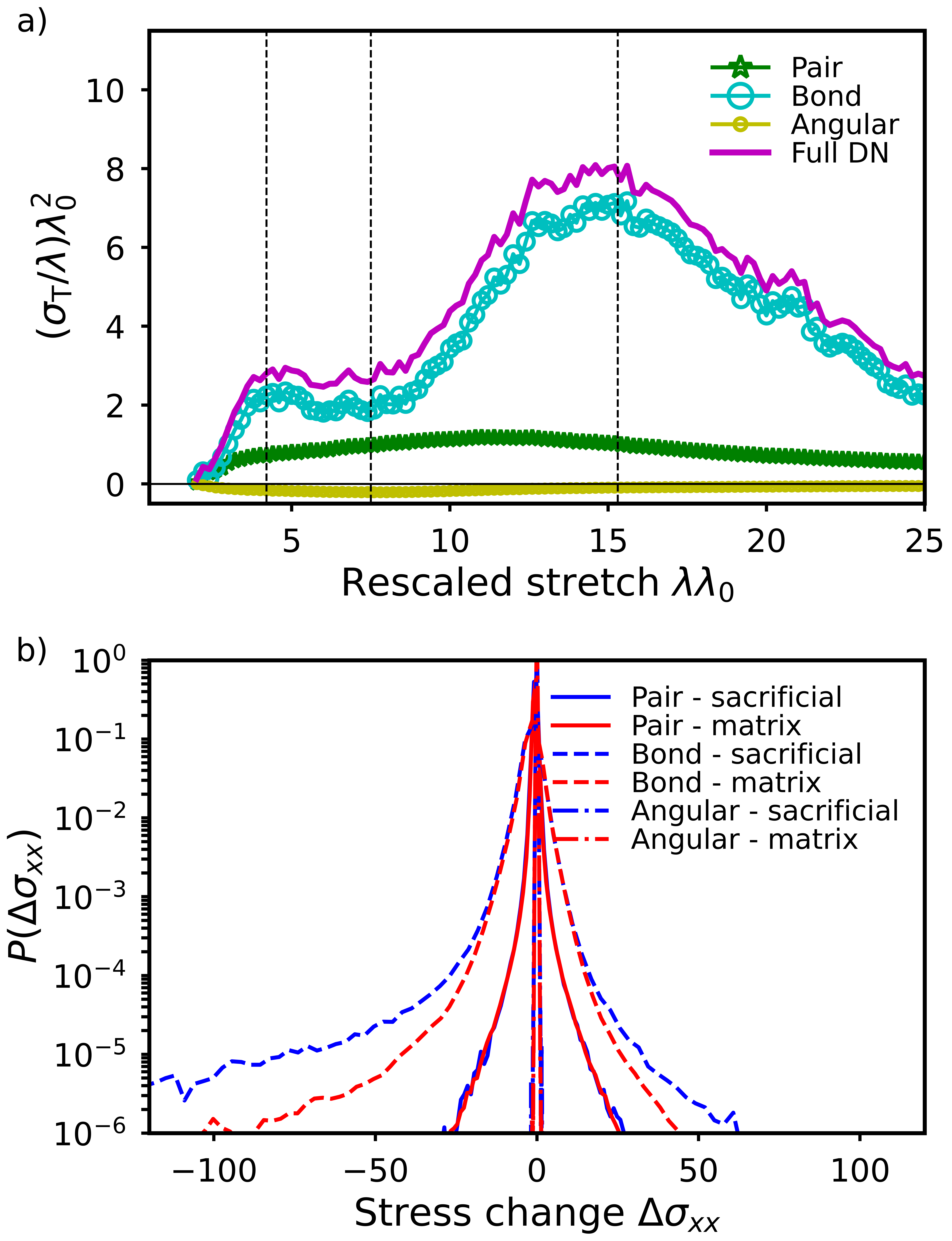}
\caption{a) Rescaled stress strain curve for the double network (magenta solid line) and for the different contributions to the stress arising from the various terms in the interaction potential: pair potential (green star symbols), bond potential (cyan big circles) and angular potential (yellow small circles). b) Distributions of the change in true stress following a bond breaking event in the sacrificial network. The different contributions to the stress are shown : pair potential term (solid line), bond potential term (dashed line) and angular potential term (dot-dashed line). The response of the two networks are shown: sacrificial (blue) and matrix (red).}
\label{fig:SI-stress_contribs_bb}
\end{figure}

\subsubsection{Noise on displacement measurements}

The displacement magnitude measured between successive snapshots always exhibits a non-zero baseline (even in the absence of bond breaking events) due to the fact that strands are not fully equilibrated.

\begin{figure}[H]
\centering
\includegraphics[width=1\columnwidth]{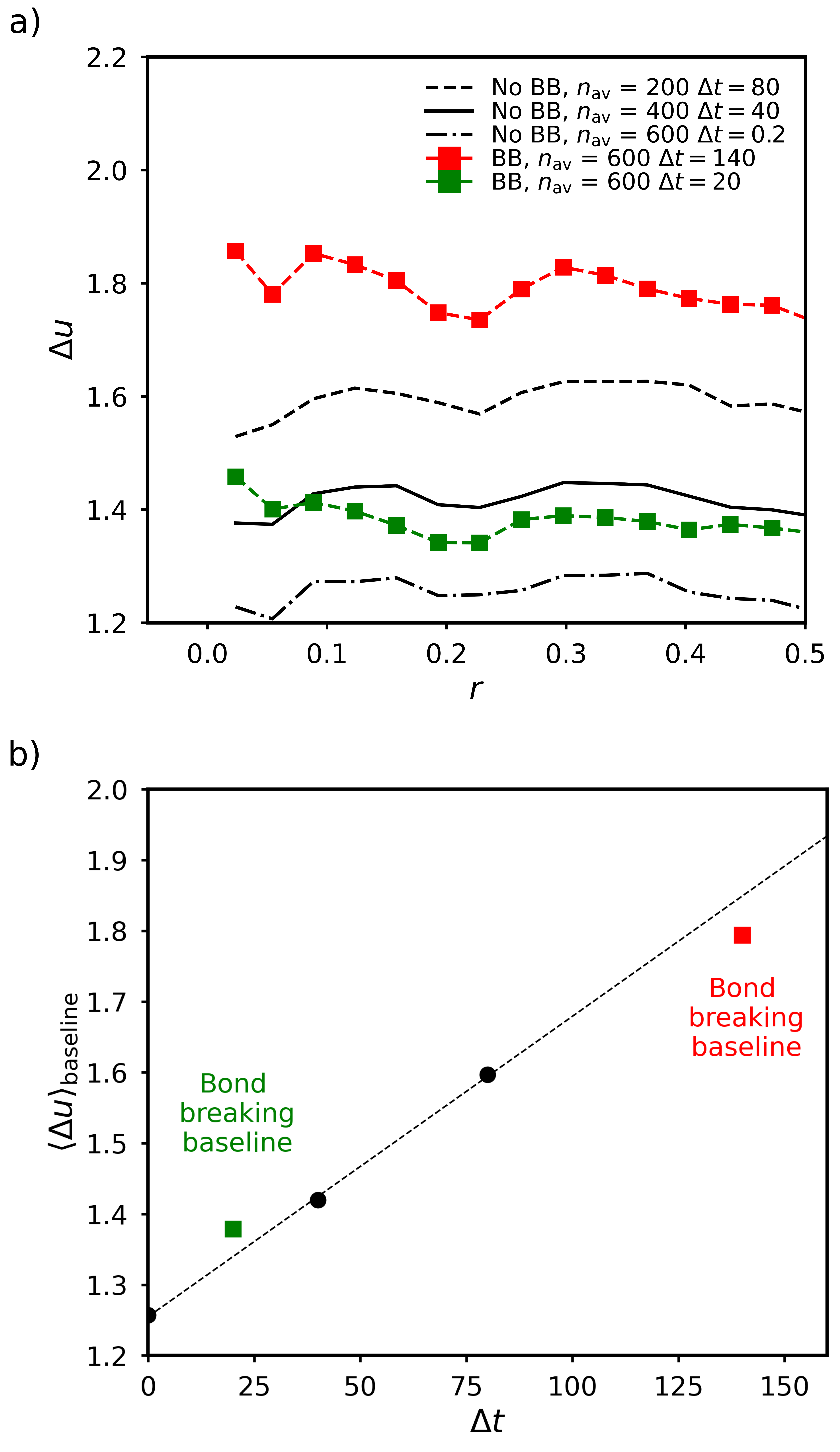}
\caption{a) Displacement magnitude $\Delta u$ as a function of $r$ (the distance to the bond breaking event in case of bond breaking, in an arbitrary frame in the absence of bond breaking) for different shift times between the averaging windows. b) Average displacement magnitude $\langle \Delta u \rangle _\mathrm{baseline}$ (measured from panel a)) as a function of the shift time $\Delta t$ in the absence of bond breaking (black dots) and baseline measured away from the bond breaking event in the case of bond breaking (average of the last 6 points in the red and green curves of panel a))}
\label{fig:SI-displacement_baseline}
\end{figure}

In Fig.~\ref{fig:SI-displacement_baseline}a, we show the displacement magnitude as a function of the distance to bond breaking (green and red squares) and in an arbitrary frame when there is no bond breaking (black lines). 


We report in Fig.~\ref{fig:SI-displacement_baseline}b the magnitude of the displacement  baseline measured from the data of panel a, far from the bond breaking event, plotting it as a function of the lag time $\Delta t$. 
We see a linear increase of the displacement as a function of lag time, as expected due to the drift of non-equilibrated polymer strands.
Note that the mean squared displacement reaches a plateau on timescales ranging between $10^3 \tau$ and $10^4 \tau$~\cite{tian2025influence}.
The data points associated with a bond breaking events occurring between $\mathrm{conf_0}$ and $\mathrm{conf_1}$ lie on the same curve.
It is therefore reasonable to assume that the displacement magnitude measured away from the bond breaking event is due to thermal drift.  We thus remove this thermal drift contribution for each bond breaking event for the displacement maps and profile shown in Fig. 9 of the main text.

\begin{acknowledgments}
We acknowledge discussions with L. Ortellado, W. C. K. Poon, L. Berthier, E. Del Gado, F. Vernerey and J. Van der Gucht. LC gratefully acknowledges support from the Institut Universitaire de France. This work was supported by the French ANR, grant n. ANR-20-CE06-0028 (MultiNet project). Numerical simulations have been performed on the GRICAD infrastructure (https://gricad.univ-grenoble-alpes.fr), which is supported by the Grenoble research communities.
\end{acknowledgments}

\nocite{*}

\bibliography{__biblio}

\end{document}